\documentclass[hyper,12pt,a4paper]{article}
\pdfoutput=1
\usepackage{mathrsfs, subfigure}
\usepackage{amsmath}
\usepackage{amsfonts}
\usepackage{amssymb}
\usepackage{graphicx}

\usepackage{jheppub}

\newcommand {\nn}    {\nonumber}
\newcommand{\be}{\begin{equation}}
\newcommand{\bea}{\begin{eqnarray}}
\newcommand{\eea}{\end{eqnarray}}
\newcommand{\ba}{\begin{array}}
\newcommand{\ea}{\end{array}}
\newcommand{\ee}{\end{equation}}

\def\bse{\begin{subequations}}
\def\ese{\end{subequations}}

\title{Holographic s-wave and p-wave Josephson junction with backreaction}
\author{Yong-Qiang Wang,}
\author{Shuai Liu}

\affiliation{Institute of Theoretical Physics, Lanzhou University,
Lanzhou 730000, People's Republic of China}
\emailAdd{yqwang@lzu.edu.cn,yuyefeng28@163.com}

\abstract{In this paper, we study the holographic models of s-wave and p-wave Josephoson junction away from probe limit
in (3+1)-dimensional spacetime, respectively. With the backreaction of the matter, we obtained the anisotropic black hole solution with the condensation of matter fields. We observe that the critical temperature of Josephoson junction decreases with increasing backreaction. In addition to this, the tunneling current and condenstion of Josephoson junction  become smaller as  backreaction grows larger, but the relationship between current and  phase difference still holds for sine function. Moreover, condenstion of Josephoson junction deceases with increasing width of junction exponentially.}

\begin{document}
\maketitle

\section{Introduction}
The AdS/CFT duality \cite{Maldacena:1997re,Maldacena:1998re,Witten:1998qj,Aharony:1999ti}, which originates from string theory, provides a novel and powerful tool to study the strongly coupled field theories in a weakly coupled gravitational system. It states a $d$-dimensional conformal field theory on the boundary is equivalent to a ($d+1$)-dimensional dual gravitational description in the bulk. Since the AdS/CFT correspondence can provide a holographically dual description of the strongly coupled system, it has received a broad range of attention. In recent years, the application of AdS/CFT correspondence to condensed matter physics has been quite successful. Especially,
 one of hot points is the study of holographic superconductor.
 In \cite{Gubser:2008px,Hartnoll:2008vx}, the authors studied the s-wave superconductor by coupling anti-de Sitter gravity to the $U(1)$ gauge field and a complex scalar field. Ones found that when  the Hawking temperature of black hole was below a critical temperature, the $U(1)$ gauge symmetry  would be broken spontaneously via the charged scalar field condensated outside the horizon. In \cite{Gubser:2008wv,Chen:2010mk,Benini:2010pr}, the authors investigated the p-wave superconductor by considering $SU(2)$ gauge field, and studied the d-wave superconductor by a symmetric, traceless second-rank tensor and a $U(1)$ gauge field in the background of the AdS black hole.  Moreover, one also studied the coexistence and competition of order parameters by holographic approach in \cite{Basu:2010fa,Musso:2013rnr,Cai:2013wma,Amoretti:2013oia,Li:2014wca,Nie:2013sda,Amado:2013lia,Nie:2014qma}. More detailed introduction of holographic superconductor can be found in \cite{Hartnoll:2009sz,Herzog:2009xv,Horowitz:2010gk,Cai:2015cya}.

In addition to the study of holographic superconductor in the probe limit, the s-wave superconductor with backreaction is also researched in \cite{Hartnoll:2008kx}. Furthermore, in \cite{Ammon:2009xh}, the authors constructed an $SU(2)$ Einstein-Yang-Mills theory with (4+1)-dimensional asymptotically anti-de Sitter charged black hole to describe p-wave superfluids with backreaction. The authors investigated the s-wave superconductor in (3+1)-dimensional with backreaction in the cases of pure Einstein and Gauss-Bonnet gravity, respectively in \cite{Brihaye:2010mr}.
The papers about holographic p-wave phase transition in Gauss-Bonnet Gravity and the s-wave superconductor in (3+1)-dimensional AdS spacetime with backreaction are presented in  \cite{Cai:2010zm} and \cite{Pan:2011ns}, respectively. There are another papers \cite{Liu:2011fy,Peng:2012vb,Pan:2012jf,Ge:2012vp,Arias:2012py,Nakonieczny:2014pma} about holographic superconductors with backreaction
 by semi-analytic and numerical computation method.
In these papers, ones find that critical temperature will become lower and condensation will become harder if the strength of backreaction becomes stronger. Furthermore, AdS/CFT correspondence has been application in the study of holographic lattice. For example, in \cite{Horowitz:2012ky}, the authors studied the optical conductivity by adding  a gravitational background lattice.  Other papers about holographic lattice can be viewed in \cite{Horowitz:2012gs,Horowitz:2013jaa,Ling:2013aya,Bao:2013ixa,Mozaffar:2013bva,Ishibashi:2013nsa,Donos:2013eha,Iizuka:2013wya,
Aprile:2014aja,Donos:2014yya,Ling:2014laa,Ling:2014bda,Chen:2015azo,Andrade:2015iyf,Alsup:2016fii}.
The DeTurck method provides a good tool for solving Einstein equations in these papers.

The model of the holographic superconductor can  also be extended to study the Josephson junction which are
associated with the experiments. As we know, the Josephson junction consists of two superconductor
materials and a weak link barrier between them   \cite{Josephson:1962zz}. The weak link can be a thin normal conductor (S-N-S) or a thin insulating barrier (S-I-S). Horowitz et al. in \cite{Horowitz:2011dz} studied the s-wave Josephson junction in probe limit by the Maxwell field coupled with a complex scalar field in  a (\(3+1\))-dimensional Schwarzschild-AdS black hole background, and observed that  the current is proportional to the sine of phase difference with AdS/CFT.
   A holographic model of  4-dimensional Josephson junction has been investigated  in \cite{Wang:2011rva,Siani:2011uj}. With the model of a designer multigravity,  the  holographic mode  of a Josephson junction array  has been constructed  in~\cite{Kiritsis:2011zq}.
 The p-wave Josephson junction was discussed by an $SU(2)$ gauge field coupled with gravity in \cite{Wang:2011ri}. In \cite{Wang:2012yj}, the authors studied (1+1)-dimensional S-I-S Josephson junction in the four-dimensional anti-de Sitter soliton background. A holographic model of superconducting quantum interference device (SQUID) was studied in \cite{Cai:2013sua,Takeuchi:2013kra}. In \cite{Li:2014xia}, authors investigated the holographic Josephson junction with Lifshitz geometry. A holographic model of hybrid and coexisting s-wave and p-wave Josephson junction was constructed in \cite{Liu:2015zca}. The authors constructed a holographic model of s-wave Josephson junction with massive gravity in \cite{Hu:2015dnl}.

The previous studies on holographic Josephson junctions are in the probe limit, however,
 it would be of great interest to further explore what role the backreaction plays in Josephson effect beyond the probe limit. For example, turning on the backreaction, we wonder that whether the current is proportional to the sine of phase difference and condensation decreases with increasing width of junction. Inspired by the previous work, we took advantage of a complex scalar field coupling to the $U(1)$ gauge field and $SU(2)$ Yang-Mills field with (3+1)-dimensional RN-AdS black hole to construct the holographic models of s-wave and p-wave Josephson junction with backreaction, respectively.

In this paper we will proceed as below. In Sect.\ref{sec2}, we set up the model of s-wave Josephson junction and analyze our numerical results. In Sect.\ref{sec3}, we write the action of the model of p-wave Josephson junction, and discuss our numerical results. We take conclusion finally in Sect.\ref{sec4}.

\section{(2+1)-Dimensional s-wave Josephson junction}\label{sec2}
\subsection{The model}
Let us begin with the Maxwell field and a charged complex scalar field in the (3+1)-dimensional Einstein gravity spacetime with a negative cosmological constant. The Lagrangian density reads
\be
\mathcal{L}=R-2\Lambda-\frac{1}{4}F^{\mu\nu}F_{\mu\nu}-|\nabla\psi-iqA\psi|^{2}-m^{2}|\psi|^{2}.\label{Lagdensity}
\ee
Here, $\Lambda=-3/\ell^{2}$ is the cosmological constant, which relates to AdS radius $\ell$. The field strength of the $U(1)$ gauge field is $F_{\mu\nu}=\partial_{\mu}A_{\nu}-\partial_{\nu}A_{\mu}$, $m$ and $q$ represent the mass and the charge of the complex scalar field $\psi$, respectively. The charge $q$ appears in the covariant derivative and controls the strength of the backreaction of the matter fields on the metric.

Since we are interested in the effect of the backreaction and to see that how it varies with the charge $q$, we have scaling transformations $A\rightarrow A/q$ and $\psi\rightarrow \psi/q$. The Lagrangian density
(\ref{Lagdensity}) changes into
\begin{align}
\mathcal{L}&=R+\frac{6}{\ell^{2}}+\kappa\mathcal{L}_{m},\\
\mathcal{L}_{m}&=-\frac{1}{4}F^{\mu\nu}F_{\mu\nu}-|\nabla\psi-iA\psi|^{2}-m^{2}|\psi|^{2},
\end{align}
where $\kappa=1/q^{2}$ is the parameter that measures the backreaction of the matter fields. From the above, we can see that the backreaction on the gravity will decrease when $q$ increases, and the large $q$ limit ($q\rightarrow \infty$) corresponds to the probe limit (non-backreaction) of the matter sources.

The equations of motion of the scalar and the electromagnetic fields which can be derived from the Lagrangian density (\ref{Lagdensity}) are as follows
\begin{align}
(\nabla_{\mu}-iA_{\mu})(\nabla^{\mu}-iA^{\mu})\psi-m^{2}\psi&=0,\label{scalarequ}\\
\nabla_{\mu}F^{\mu\nu}-i[\psi^{\ast}(\nabla^{\nu}-iA^{\nu})\psi-\psi(\nabla^{\nu}+iA^{\nu})\psi^{\ast}]&=0,\label{maxwellequ}
\end{align}
and Einstein equations
\begin{align}
R_{\mu\nu}+\frac{3}{\ell^{2}}g_{\mu\nu}-\kappa(\frac{1}{2}F_{\mu\lambda}{F_{\nu}}^{\lambda}-\frac{1}{8}F_{\lambda\delta}F^{\lambda\delta}g_{\mu\nu}+
\frac{1}{2}m^{2}|\psi|^{2}g_{\mu\nu}\nn\\
+\frac{1}{2}[(\nabla_{\mu}\psi-i A_{\mu}\psi)(\nabla_{\nu}\psi^{\ast}+i A_{\nu}\psi^{\ast})+\mu\leftrightarrow\nu])=0.\label{einsteinequ}
\end{align}

For the charged scalar field $\psi=0$, the solution of Einstein equations (\ref{einsteinequ}) is the well-known Reissner-Nordstr\"{o}m-AdS (RN-AdS) black hole. The solution with a spherically symmetric can be written as follows
\be
ds^{2}=\frac{\ell^{2}}{z^{2}}[(z-1)H(z)dt^{2}+\frac{dz^{2}}{(1-z)H(z)}+dx^{2}+dy^{2}],
\ee
where $H(z)=1+z+z^{2}-\frac{\kappa}{4}\mu_{0}^{2}z^{3}$, and $\mu_{0}$ is the chemical potential for $U(1)$ charge. The horizon of black hole is at $z=1$ and the boundary of asymptotical AdS spacetime is at $z=0$. The Hawking temperature, which can be regarded as the temperature of the holographic superconductors, is given by
\be
T=\frac{(12-\kappa\mu_{0}^{2})}{16\pi\ell} .
\ee

It is well known that there is a critical temperature $T_{c}$. When $T=T_{c}$, a charged scalar condensation begins to occur. For $T<T_{c}$, the black hole will have a scalar hair with breaking the $U(1)$ gauge symmetry spontaneously and brings about superconducting phenomena of the (2+1)-dimensional dual theory on the boundary. As for $T>T_{c}$, the black hole with a scalar hair degrades into RN-AdS black hole.
\subsection{The ansatz and asymptotic forms}
In order to build a holographic model of the s-wave Josephson junction, we introduce two bulk gauge fields  $A_{z}$ and $A_{x}$ on the basis of the model of s-wave superconductor. The gauge field $A_{x}$ is related to the non-vanished current on the boundary. In addition, the matter fields must depend on spatial coordinates.

Considering the above reasons, an ansatz of matter fields should be  described  as below
\be
\psi=|\psi|e^{i\phi},~~A=A_{t}dt+A_{z}dz+A_{x}dx,\label{matteransatz}
\ee
where $|\psi|$, $\phi$, $A_{t}$, $A_{z}$ and $A_{x}$ all depend upon the spatial coordinate $x$ and the radial
coordinate $z$. Furthermore, we take the metric ansatz as
\begin{align}
ds^{2}=\frac{\ell^{2}}{z^{2}}[(z-1)H(z)E_{1}(dt+E_{7}dx)^{2}+\frac{E_{2}(dz+(1-z)H(z)E_{6}dt)^{2}}{(1-z)H(z)}\nn\\
+E_{3}(dx+E_{5}dz)^{2}+E_{4}dy^{2}],\label{metricansatz}
\end{align}
where $E_{i}~(i=1,2,3,4,5,6,7)$ are seven functions of $z$ and $x$. Note that the metric may be a non-diagonal one, in which the non-diagonal function $E_{5}$ is required due to the $x$ dependent spatial coordinate, and the functions $E_{6}$ and $E_{7}$ are associated with non-vanishing $A_{z}$ and $A_{x}$, respectively. Thus the holographic model of s-wave Josephson junction would be along the $x$ direction. The function $\psi$ can be taken to be real by introducing  the new $U(1)$ gauge invariance  $M_{\mu}\equiv A_{\mu}-\partial_{\mu}\phi$.

The scalar field, Maxwell and Einstein equations of motion with the ansatzs (\ref{matteransatz}) and (\ref{metricansatz}) are a set of non-linear coupled partial differential equations. Seven equations which come from the Einstein equations (\ref{einsteinequ}) are second-order PDEs with respect to $E_{i}$. Two equations from Eq.(\ref{scalarequ}), one is a second-order PDE and the other one is a first-order PDE, are called as constraint equations. The remaining three equations which are from the Eqs.(\ref{maxwellequ}) are also second-order PDEs. It is not convenient to write down all of the twelve equations in our paper, for each equation contains hundreds or thousands of terms. It is obvious that we should solve them numerically instead of seeking the analytical solutions. Before numerical program, we should obtain the asymptotic behaviors of $|\psi|$, $M_{t}$, $M_{z}$, $M_{x}$ and $E_{i}$ at $z=0$. To know these asymptotic forms is equivalent to know the boundary conditions we need.

On the AdS boundary, the asymptotic behaviors of the functions $E_{i}$ take the following forms
\begin{subequations}
\begin{align}
E_{1}(z,x)&\rightarrow 1+\mathcal{O}(z^{3}),\label{e1}\\
E_{2}(z,x)&\rightarrow 1+\mathcal{O}(z^{3}),\\
E_{3}(z,x)&\rightarrow 1+\mathcal{O}(z^{3}),\\
E_{4}(z,x)&\rightarrow 1+\mathcal{O}(z^{3}),\\
E_{5}(z,x)&\rightarrow \mathcal{O}(z^{4}),\\
E_{6}(z,x)&\rightarrow \mathcal{O}(z^{4}),\\
E_{7}(z,x)&\rightarrow \mathcal{O}(z^{4}).\label{e7}
\end{align}
\end{subequations}
Obviously, when $z$ goes to zero, the metric (\ref{metricansatz}) will approach to the Reissner-Nordstr\"{o}m-AdS metric with the above asymptotic forms.

Near the boundary $z=0$, the scalar field and Maxwell field have the following asymptotic forms
\begin{subequations}
\begin{align}
|\psi|(z,x)&\rightarrow z^{\Delta_{-}}\psi^{(-)}(x)+z^{\Delta_{+}}\psi^{(+)}(x)+\mathcal{O}(z^{1+\Delta_{+}}),\\
M_{t}(z,x)&\rightarrow \mu(x)-z\rho(x)+\mathcal{O}(z^{2}),\\
M_{z}(z,x)&\rightarrow \mathcal{O}(z),\\
M_{x}(z,x)&\rightarrow \nu(x)+Jz+\mathcal{O}(z^{2}),
\end{align}
\end{subequations}
with
\be
\Delta_{\pm}=\frac{(3\pm\sqrt{9+4m^{2}})}{2},
\ee
where $m$ is the mass of the scalar field and the values of $m^{2}$ must satisfy the Breitenlohner-Freedman (BF) bound $m^{2}\geq-9/4$ \cite{Breitenlohner:1982bm} for the (3+1)-dimensional spacetime. According to AdS/CFT duality, $\psi^{(\pm)}(x)$ are the corresponding expectation values of the dual scalar operators $\langle\mathcal{O_{\pm}}\rangle$, respectively. In this paper, we will set $\psi^{(-)}=0$ and take $\langle\mathcal{O}\rangle=\langle\mathcal{O}_{+}\rangle=\psi^{(+)}$ to describe the scalar condensation. The coefficients $\mu(x)$, $\rho(x)$, $\nu(x)$ and $J$ are the chemical potential, charge density, the velocity of superfluid and the constant current in the dual field theory, respectively \cite{Basu:2008st,Herzog:2008he,Arean:2010xd,Sonner:2010yx,Horowitz:2008bn,Arean:2010zw,Zeng:2010fs,Arean:2010wu}.

In order to describe a Josephson junction, we still adopt the chemical potential $\mu(x)$ in \cite{Horowitz:2011dz} as follows
\be
\mu(x)=\mu(\infty)\left\{1-\frac{1-\epsilon}{2\tanh(\frac{L}{2\sigma})}\left[\tanh\left(\frac{x+\frac{L}{2}}{\sigma}\right)-\tanh\left(\frac{x-\frac{L}{2}}{\sigma}\right)\right]\right\},\label{chempoten}
\ee
where the chemical potential $\mu(x)$ is proportional to $\mu(\infty)\equiv\mu(+\infty)=\mu(-\infty)$ at $x=\pm\infty$, and $L$ is the width of junction. The parameters $\epsilon$ and $\sigma$ control the steepness and depth of junction, respectively.

\begin{figure}
  \begin{center}
  \includegraphics[width=0.9\textwidth,height=0.4\textheight]{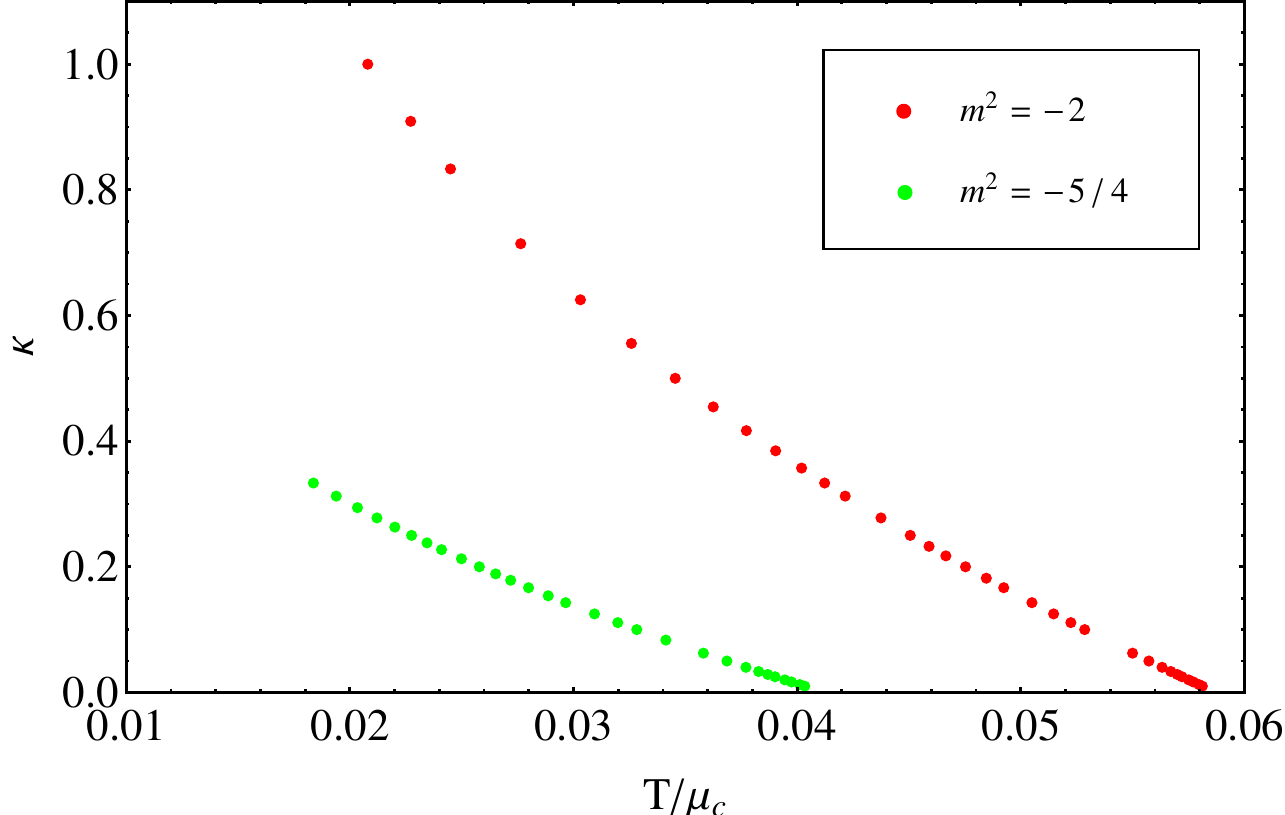}
  \end{center}
  \caption{The plot shows that the temperature $T/\mu_{c}$ decays as the strength of backreaction $\kappa$ increases with fixing the value of $m^{2}$. We take two values of $m^{2}$=$-2$(red), $-5/4$(green). When $m^{2}$ grows, $T/\mu_{c}$ will decrease.}
\label{kappaT}
\end{figure}

Next, we introduce the critical temperature $T_{c}$ of the Josephson junction, which is proportional to $\mu(\infty)$
\be
T_{c}=T\frac{\mu(\infty)}{\mu_{c}},
\ee
where $\mu_{c}$ is the critical chemical potential of the superconductor. In figure \ref{kappaT}, we plot the relationship between the temperature $T/\mu_{c}$ and the strength of backreaction $\kappa$ with $m^{2}=-2,-5/4$. It indicates that $T/\mu_{c}$ decays as $\kappa$ increases with fixing the value of $m^{2}$, and when $m^{2}$ grows, $T/\mu_{c}$ will decrease.

When $m^{2}=-2$,
\begin{align}
T_{c}&\approx0.0588\mu(\infty),\quad \quad \kappa=0,\label{Tsc1}\\
T_{c}&\approx0.0581\mu(\infty),\quad \quad \kappa=1/100,\\
T_{c}&\approx0.0550\mu(\infty),\quad \quad \kappa=1/16,\\
T_{c}&\approx0.0505\mu(\infty),\quad \quad \kappa=1/7.
\end{align}
When $m^{2}=-5/4$,
\begin{align}
T_{c}&\approx0.0413\mu(\infty),\quad \quad \kappa=0,\\
T_{c}&\approx0.0403\mu(\infty),\quad \quad \kappa=1/100,\\
T_{c}&\approx0.0358\mu(\infty),\quad \quad \kappa=1/16,\\
T_{c}&\approx0.0296\mu(\infty),\quad \quad \kappa=1/7\label{Tsc8}.
\end{align}
From Eqs.($\ref{Tsc1}$)-($\ref{Tsc8}$), we can see that for the value of $m^{2}$ is fixed, the critical temperature $T_{c}$ drops with increasing strength of backreaction $\kappa$, and $T_{c}$ decreases with increasing $m^{2}$ by fixing $\kappa$.

In condensed matter physics, the current of Josephson junction has to be gauge invariant because of the presence of a gauge field $A_{\mu}$. So, we need to consider the gauge invariant phase difference. Thus, in our holographic model case, the gauge invariant phase difference $\gamma$ across the junction can be defined as
\be
\gamma=-\int_{-\infty}^{+\infty}dx[\nu(x)-\nu(\pm\infty)].\label{phase}
\ee

\subsection{The numerical results}
\begin{figure}
\centering
\subfigure[]{\includegraphics[width=0.49\linewidth]{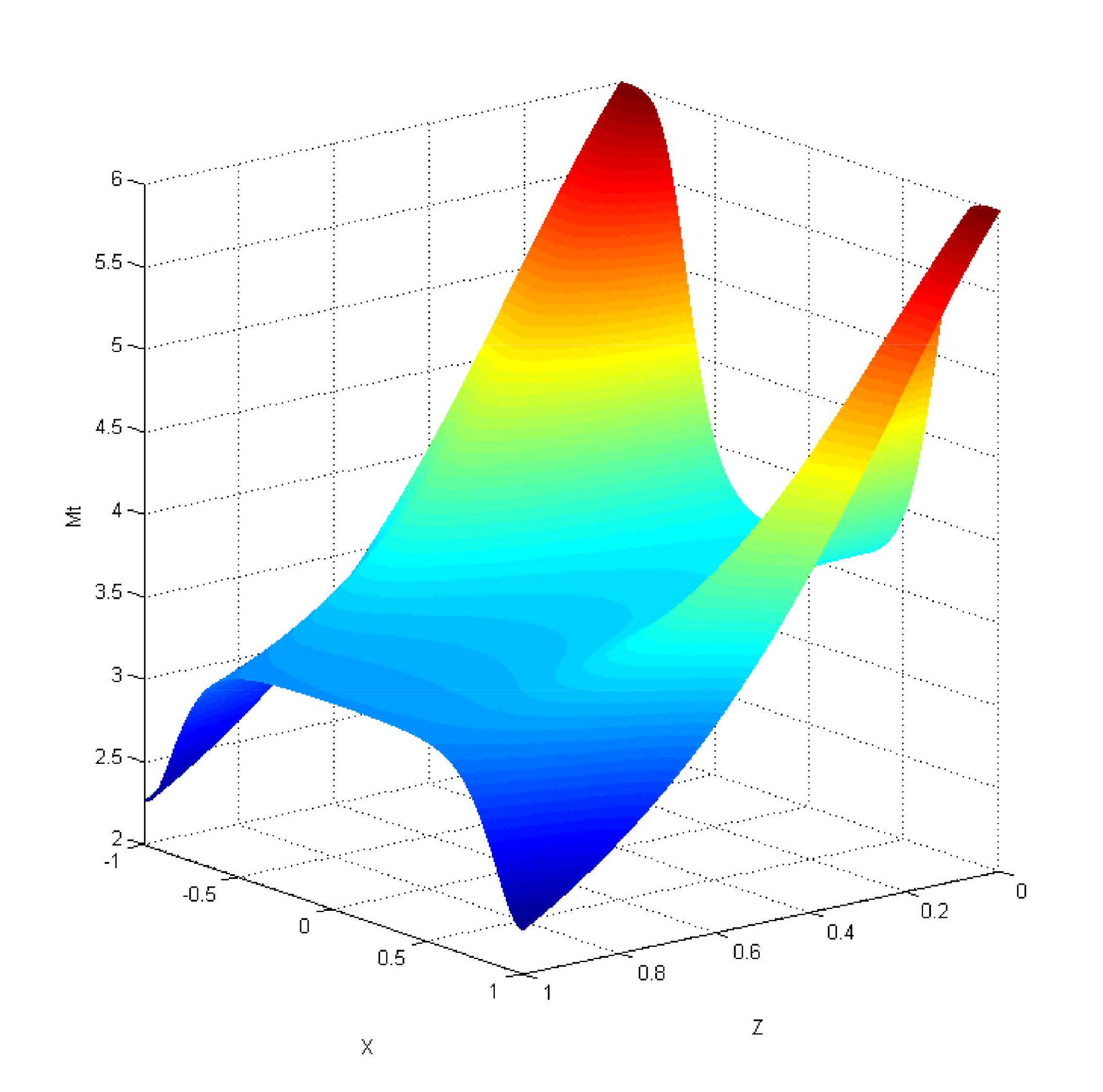}}
\hfill
\subfigure[]{\includegraphics[width=0.49\linewidth]{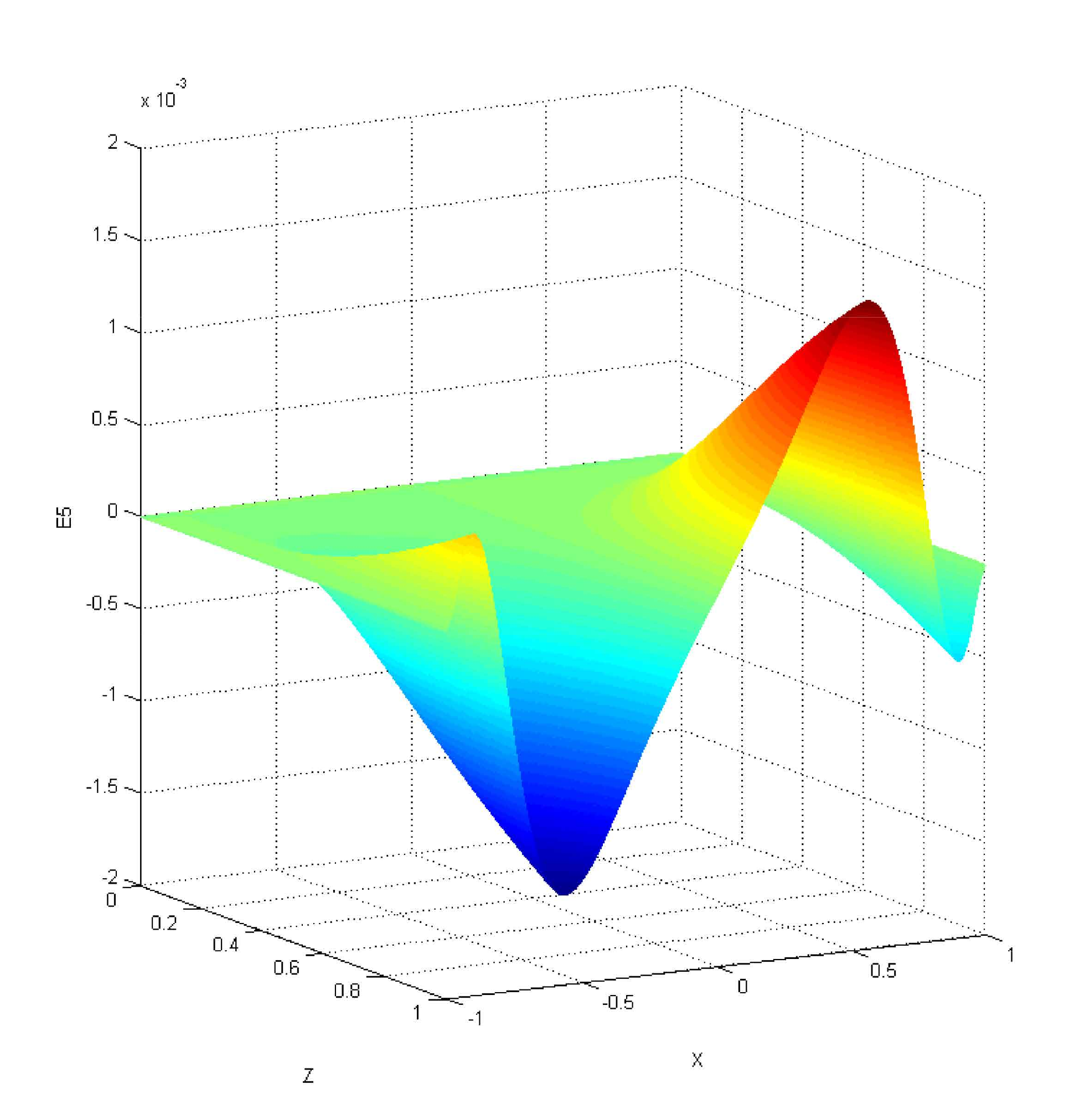}}
\caption{(a) The profile of $M_{t}$. (b) The profile of $E_{5}$. In all the plots, the parameters are $m^{2}=-2$, $\mu0=1$, $\mu_{\infty}=6$, $L=3$, $\epsilon=0.6$, $\sigma=0.5$ and $\kappa=1/100$.}
\label{swaveMtE5}
\end{figure}

For completeness of our study, we carry on numerical computation in this subsection. To make our work easier, we choose $m^{2}=-2$. We show the profiles of the functions $M_{t}$ and $E_{5}$ in figure $\ref{swaveMtE5}$, respectively. We can see that $M_{t}$ is even, and $E_{5}$ is odd.

\begin{figure}
  \begin{center}
  \includegraphics[width=0.9\textwidth,height=0.4\textheight]{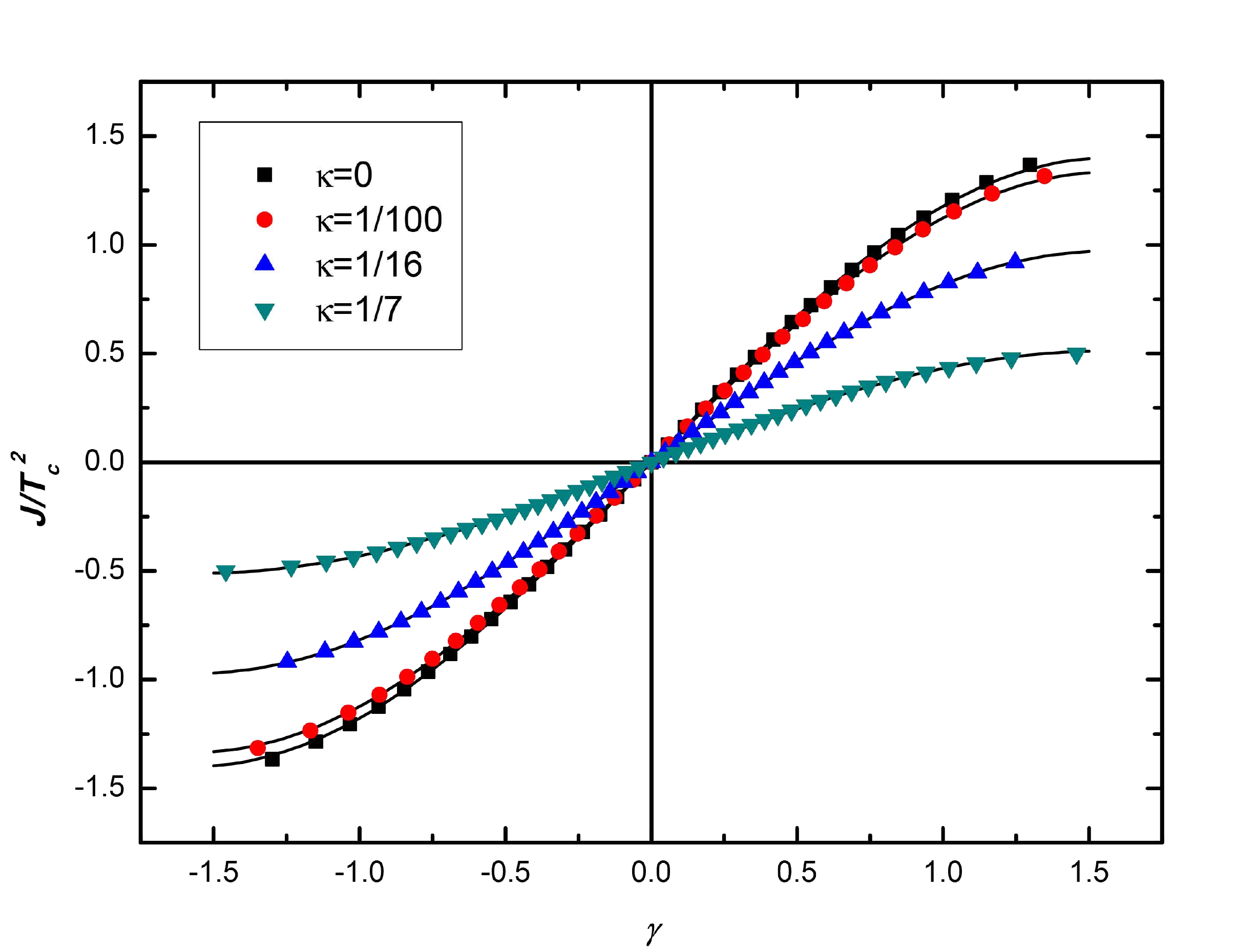}
  \end{center}
  \caption{(Color online) We plot the current $J/T_{c}^{2}$ as a function of the  phase difference $\gamma$. Curves from top to bottom are for choosing $\kappa=0$, $\kappa=1/100$, $\kappa=1/16$ and $\kappa=1/7$. The parameters are $m^{2}=-2$, $\mu0=1$, $\mu_{\infty}=6$, $L=3$, $\epsilon=0.6$ and $\sigma=0.5$.}
\label{Jgammam=-2k}
\end{figure}

For the increasing strength of backreaction, the relationship between the current $J/T_{c}^{2}$ and the phase difference $\gamma$ as shown in figure $\ref{Jgammam=-2k}$. We take the strength of backreaction $\kappa=0,1/100,1/16,1/7$. The fitting results are as follows
\begin{align}
J/T_{c}^{2}&\approx1.39981sin\gamma,\quad \quad \kappa=0,\\
J/T_{c}^{2}&\approx1.33521sin\gamma,\quad \quad \kappa=1/100,\\
J/T_{c}^{2}&\approx0.97168sin\gamma,\quad \quad \kappa=1/16,\\
J/T_{c}^{2}&\approx0.51218sin\gamma,\quad \quad \kappa=1/7.
\end{align}
To our surprise, the current $J/T_{c}^{2}$ is still proportional to the sine of the phase difference $\gamma$ with backreaction($\kappa\neq0$). Moreover, the amplitude of $J/T_{c}^{2}$ goes down as $\kappa$ increases.

\begin{figure}
\centering
\subfigure[]{\includegraphics[width=0.49\linewidth]{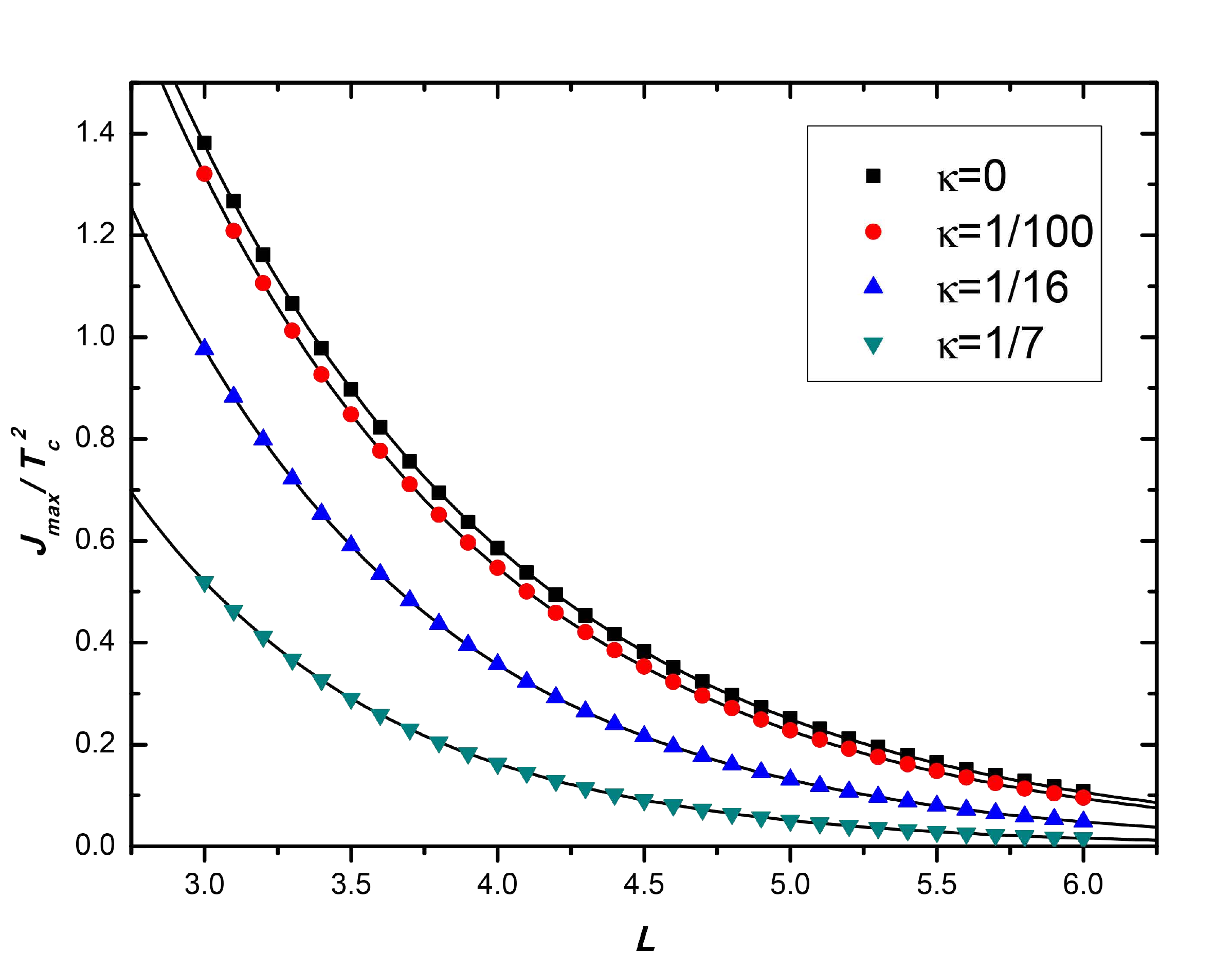}}
\hfill
\subfigure[]{\includegraphics[width=0.49\linewidth]{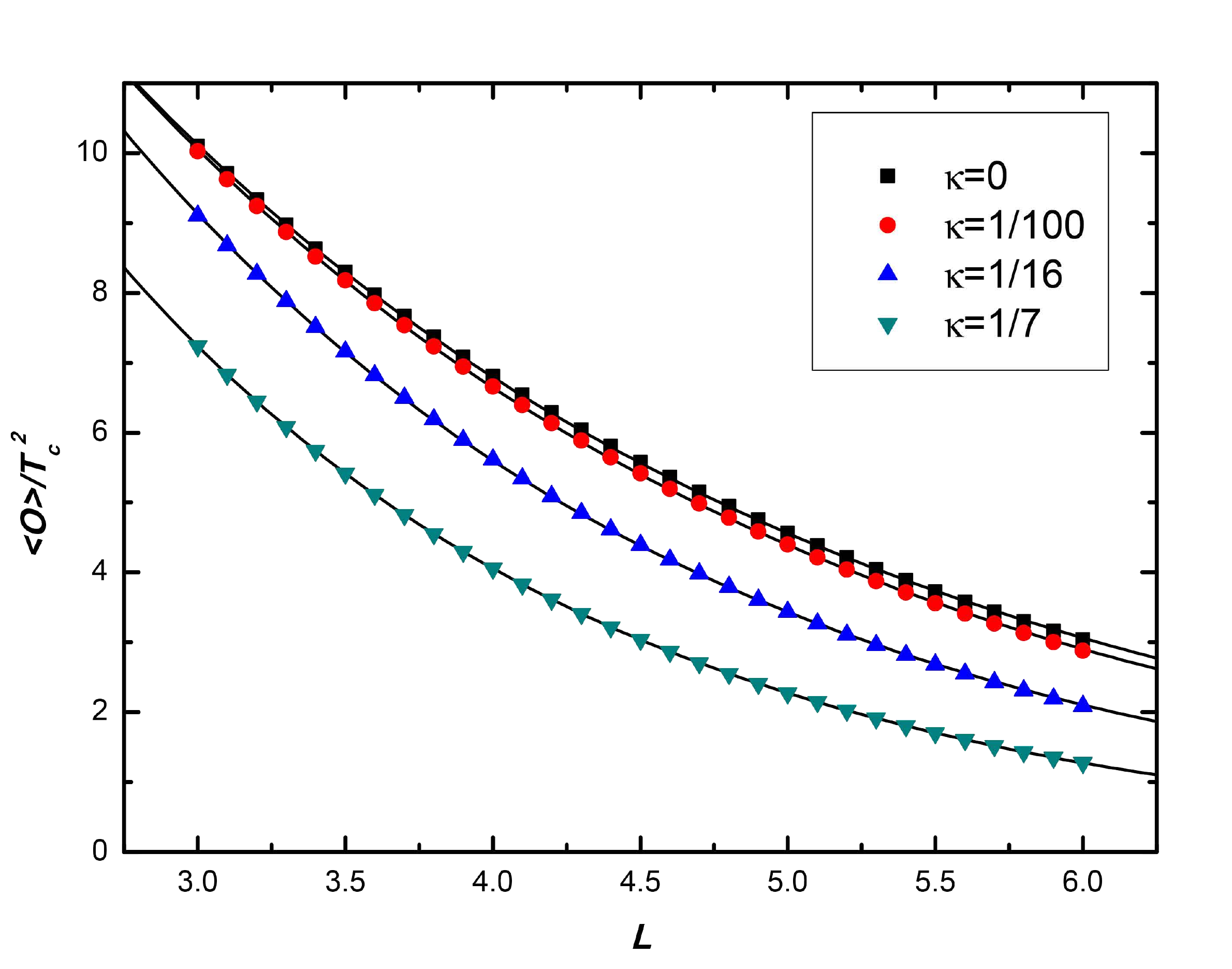}}
\caption{(a) The maximal current $J_{max}/T_{c}^{2}$ versus the width of junction $L$ for $m^{2}=-2$. For the fixing $\kappa$, $J_{max}/T_{c}^{2}$ varies with increasing $L$. (b) The condensation $\langle O\rangle/T_{c}^{2}$ versus the width of junction $L$ for $m^{2}=-2$. For the fixing $\kappa$, $\langle O\rangle/T_{c}^{2}$ varies as a function of $L$. Curves from top to bottom are for choosing $\kappa=0$, $\kappa=1/100$, $\kappa=1/16$ and $\kappa=1/7$. The curves fit exponential curves well. In all the plots, the parameters are $\mu0=1$, $\mu_{\infty}=6$, $\epsilon=0.6$ and $\sigma=0.5$.}
\label{JmaxLOLm=-2k}
\end{figure}

The relationship between the maximal current $J_{max}/T_{c}^{2}$ and the width of the junction $L$ with increasing strength of backreaction $\kappa=0,1/100,1/16,1/7$ in the figure $\ref{JmaxLOLm=-2k}$ (a). The fitting results are as follows
\begin{align}
J_{max}/T_{c}^{2}&\approx17.8455e^{-L/1.17112},\quad \quad \kappa=0,\\
J_{max}/T_{c}^{2}&\approx18.4674e^{-L/1.13660},\quad \quad \kappa=1/100,\\
J_{max}/T_{c}^{2}&\approx19.9195e^{-L/0.994678},\quad \quad \kappa=1/16,\\
J_{max}/T_{c}^{2}&\approx17.0124e^{-L/0.860052},\quad \quad \kappa=1/7.
\end{align}
It is surprising that when $\kappa$ is fixed, $J_{max}/T_{c}^{2}$  will still decrease exponentially with increasing $L$. In addition, $J_{max}/T_{c}^{2}$ becomes smaller with increasing $\kappa$. Figure $\ref{JmaxLOLm=-2k}$ (b) shows the relationship between the condensation $\langle O\rangle/T_{c}^{2}$ and the width of junction $L$. The fitting results are as below
\begin{align}
\langle O\rangle/T_{c}^{2}&\approx33.4411e^{-L/(2\times1.25493)},\quad \quad \kappa=0,\\
\langle O\rangle/T_{c}^{2}&\approx34.8072e^{-L/(2\times1.20773)},\quad \quad \kappa=1/100,\\
\langle O\rangle/T_{c}^{2}&\approx39.5984e^{-L/(2\times1.02245)},\quad \quad \kappa=1/16,\\
\langle O\rangle/T_{c}^{2}&\approx41.0873e^{-L/(2\times0.863886)},\quad \quad \kappa=1/7.
\end{align}
These indicates that $\langle O\rangle/T_{c}^{2}$  will still decrease exponentially with increasing $L$, when $\kappa$ is fixed. In addition, $\langle O\rangle/T_{c}^{2}$ goes down as $\kappa$ increases.

\begin{figure}
  \begin{center}
  \includegraphics[width=0.9\textwidth,height=0.4\textheight]{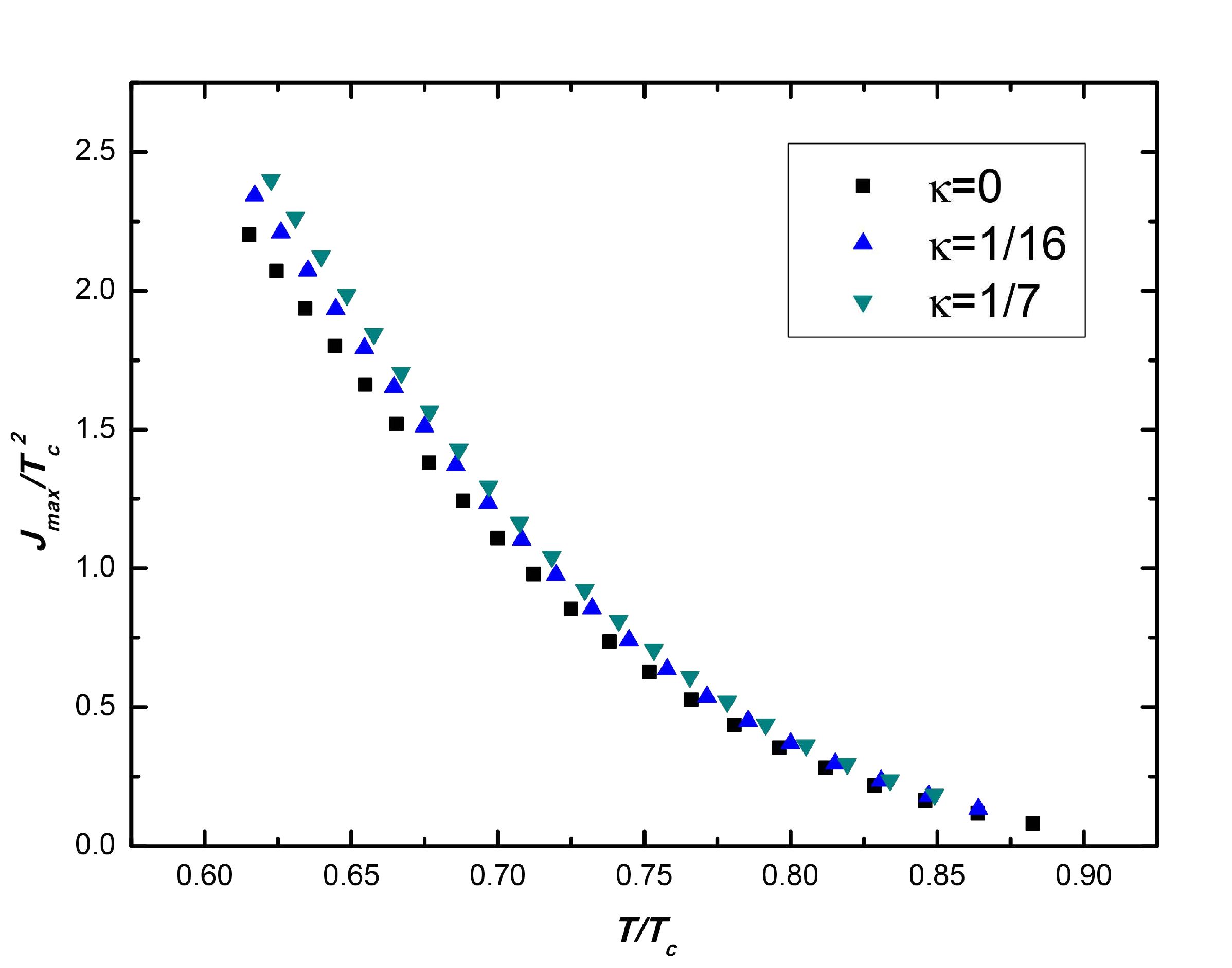}
  \end{center}
  \caption{The plot shows that $J_{max}/T_{c}^{2}$ decreases with increasing $T/T_{c}$ by fixing $\kappa$, and $J_{max}/T_{c}^{2}$ will go up, when $\kappa$ increases. We use $m^{2}=-2$, $\mu0=1$, $\mu_{\infty}=6$, $L=3$, $\epsilon=0.6$ and $\sigma=0.5$. From top to bottom, curves correspond to $\kappa=1/7,1/16,0$.}
\label{JmaxTm=-2k}
\end{figure}

The relationship between the maximal current $J_{max}/T_{c}^{2}$ and the temperature $T/T_{c}$ with the backreaction $\kappa=1/7,1/16,0$ is in figure $\ref{JmaxTm=-2k}$. From the figure, we can see that $J_{max}/T_{c}^{2}$ decreases with increasing $T/T_{c}$ by fixing $\kappa$. It is noteworthy that $J_{max}/T_{c}^{2}$ will go up, when $\kappa$ increases.

From the above results we find that the backreaction can hinder the generation of condensation and current, and make the critical temperature lower. In essence, when the intensity of backreaction $\kappa$ becomes larger,   the Cooper pairs generate harder, the charge $q$ ($\kappa=1/q^2$) and the coherence length become smaller. Thus these factors lead to the above numerical results. For the $J_{max}/T^{2}_{c}$ which varies with $T/T_{c}$ grows with increasing $\kappa$, it makes sense that stronger backreaction weakens condensation and lower temperature enhances it, but the latter plays a dominate role in this competition.

\begin{figure}
  \begin{center}
  \includegraphics[width=0.9\textwidth,height=0.4\textheight]{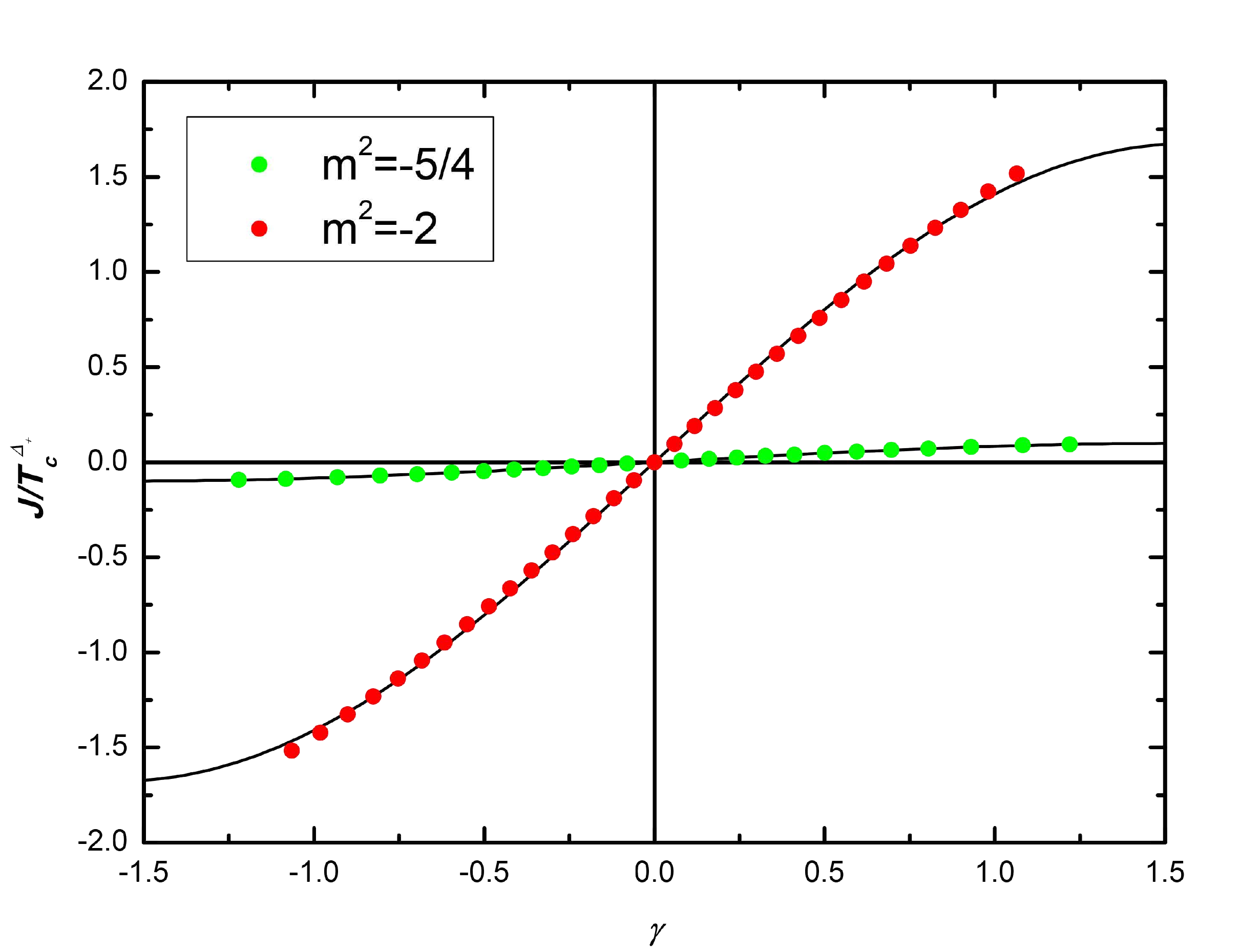}
  \end{center}
  \caption{The current $J/T^{\Delta_{+}}_{c}$ is proportional to the sine of $\gamma$ and goes down as $m^{2}$ increases. The value of $m^{2}=-2$(red), $-5/4$(green). The parameters are $\mu0=1$, $\mu_{\infty}=7.5$, $L=3$, $\epsilon=0.5$, $\sigma=0.5$ and $\kappa=1/100$.}
\label{Jgammam}
\end{figure}

\begin{figure}
\centering
\subfigure[]{\includegraphics[width=0.49\linewidth]{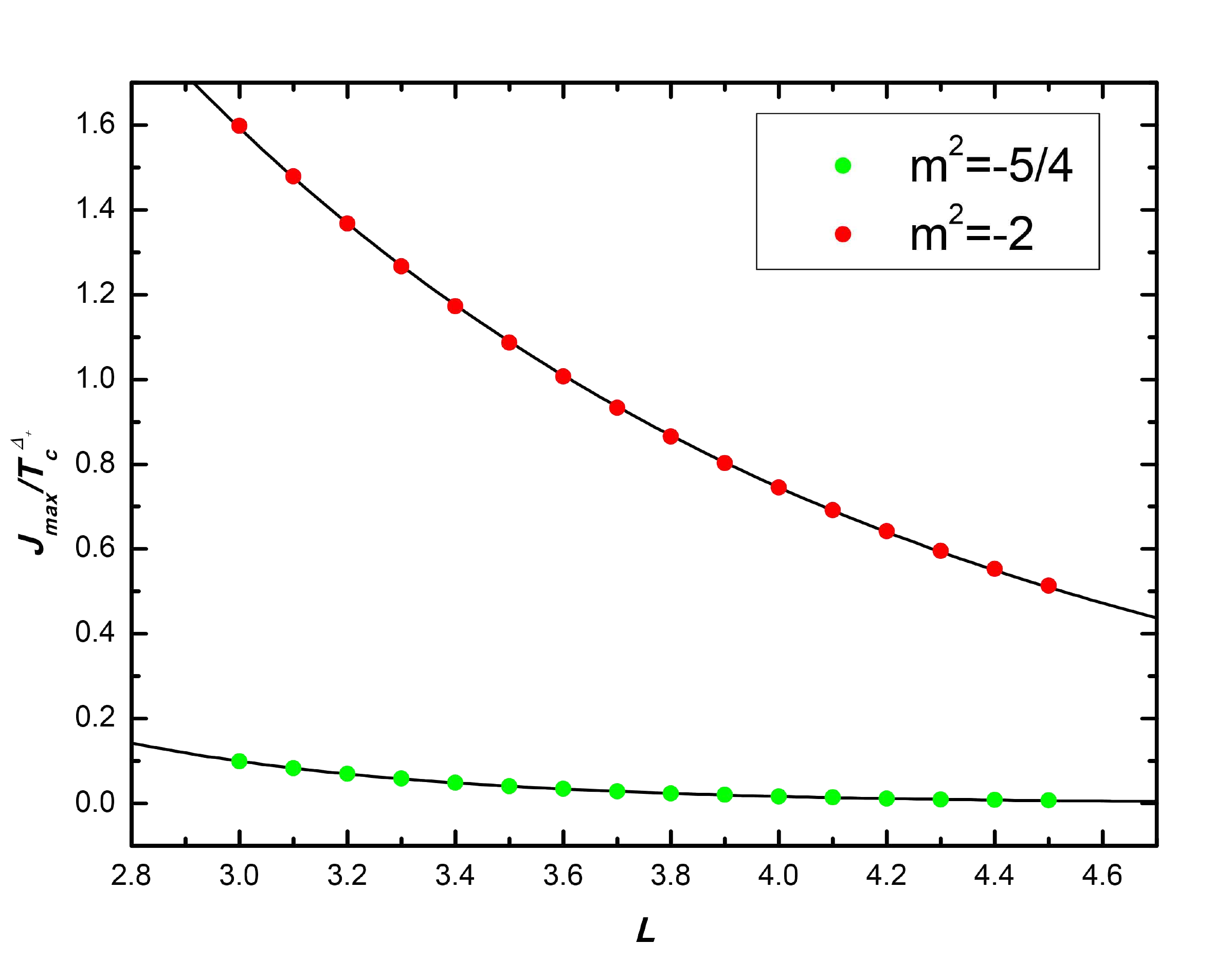}}
\hfill
\subfigure[]{\includegraphics[width=0.49\linewidth]{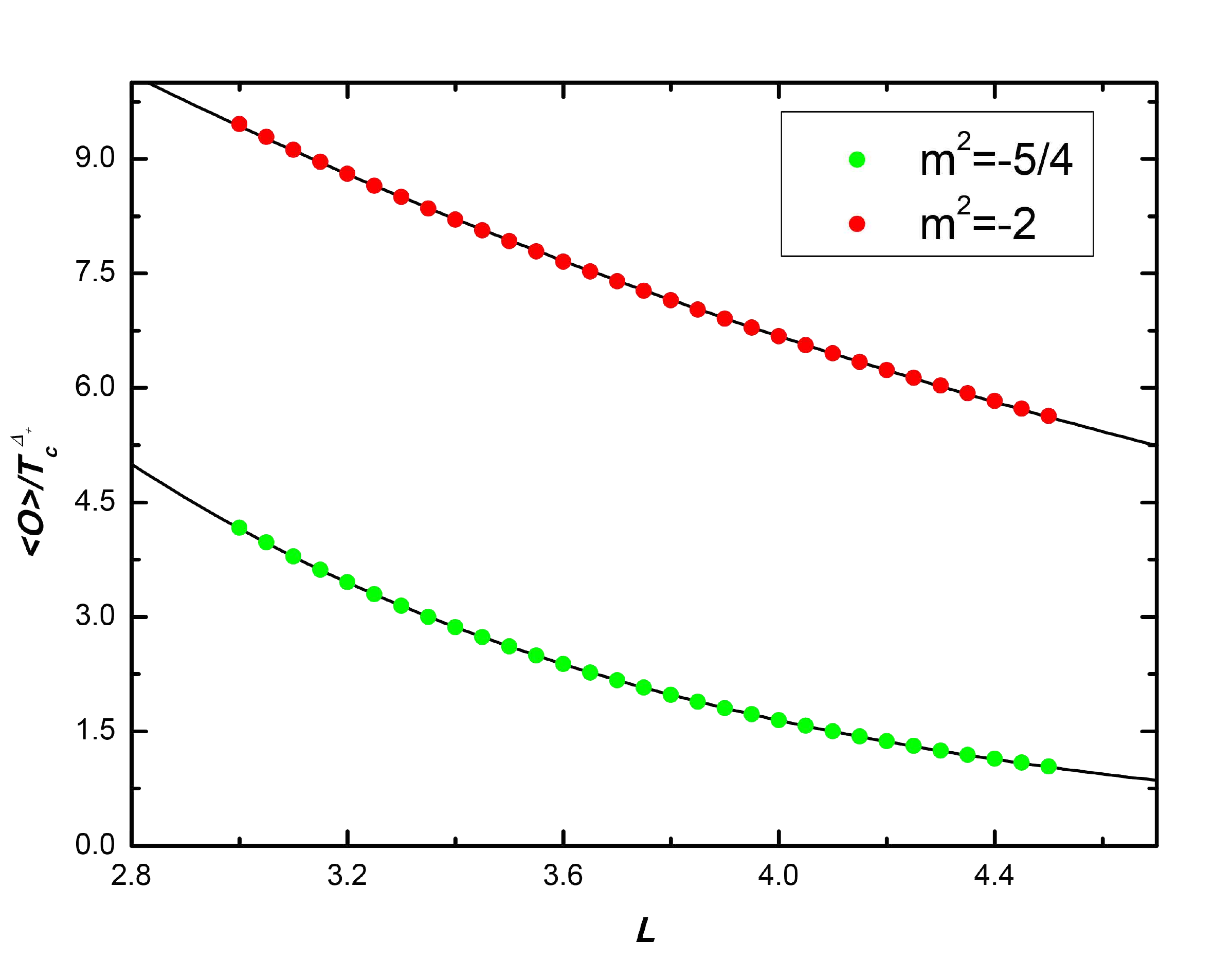}}
\caption{(a) The maximal current $J_{max}/T^{\Delta_{+}}_{c}$ decreases exponentially with growing $L$. (b) The condensation $\langle O\rangle/T^{\Delta_{+}}_{c}$ decreases exponentially with growing $L$. When $m^{2}$ increases, $J_{max}/T^{\Delta_{+}}_{c}$ and $\langle O\rangle/T^{\Delta_{+}}_{c}$ will decrease, respectively. The value of $m^{2}=-2$(red), $-5/4$(green). The parameters $\mu0=1$, $\mu_{\infty}=6$, $\epsilon=0.5$, $\sigma=0.5$ and $\kappa=1/100$.}
\label{JmaxLOLm}
\end{figure}

In the next moment, we take the value of $m^{2}=-5/4$ to compare with the case of $m^{2}=-2$ by $\kappa=1/100$. In figure $\ref{Jgammam}$, we represent the relationship between the current $J/T^{\Delta_{+}}_{c}$ and the phase difference $\gamma$ with different value of $m^{2}$. The fitting results are shown as below
\begin{align}
J/T_{c}^{\Delta_{+}}\approx\left\{ \begin{array}{ll}
 1.6758sin\gamma & \qquad\textrm{for \quad$m^{2}=-2$},\\
 0.0993sin\gamma & \qquad\textrm{for \quad$m^{2}=-5/4$}.
 \end{array} \right.
\end{align}
From the above, we see that for the fixing $\kappa$, the current goes down as $m^{2}$ increases.
Figure $\ref{JmaxLOLm}$ (a) and (b) show that the maximal current $J_{max}/T^{\Delta_{+}}_{c}$ and the condensation $\langle O\rangle/T^{\Delta_{+}}_{c}$ decrease with increasing $L$, respectively. The fitting results are
\begin{align}
J_{max}/T_{c}^{\Delta_{+}}\approx\left\{ \begin{array}{ll}
 15.5660e^{-L/1.3160} & \qquad\textrm{for \quad$m^{2}=-2$},\\
 21.7297e^{-L/0.5570} & \qquad\textrm{for \quad$m^{2}=-5/4$}.
 \end{array} \right.
\end{align}
and
\begin{align}
\langle O\rangle/T_{c}^{\Delta_{+}}\approx\left\{ \begin{array}{ll}
 26.5676e^{-L/(2\times1.448)} & \qquad\textrm{for \quad$m^{2}=-2$},\\
 67.2095e^{-L/(2\times0.5360)} & \qquad\textrm{for \quad$m^{2}=-5/4$}.
 \end{array} \right.
\end{align}
From the above, we can see that when $m^{2}$ increases, $J_{max}/T^{\Delta_{+}}_{c}$ and $\langle O\rangle/T^{\Delta_{+}}_{c}$ will decrease, respectively.

\begin{figure}
  \begin{center}
  \includegraphics[width=0.9\textwidth,height=0.4\textheight]{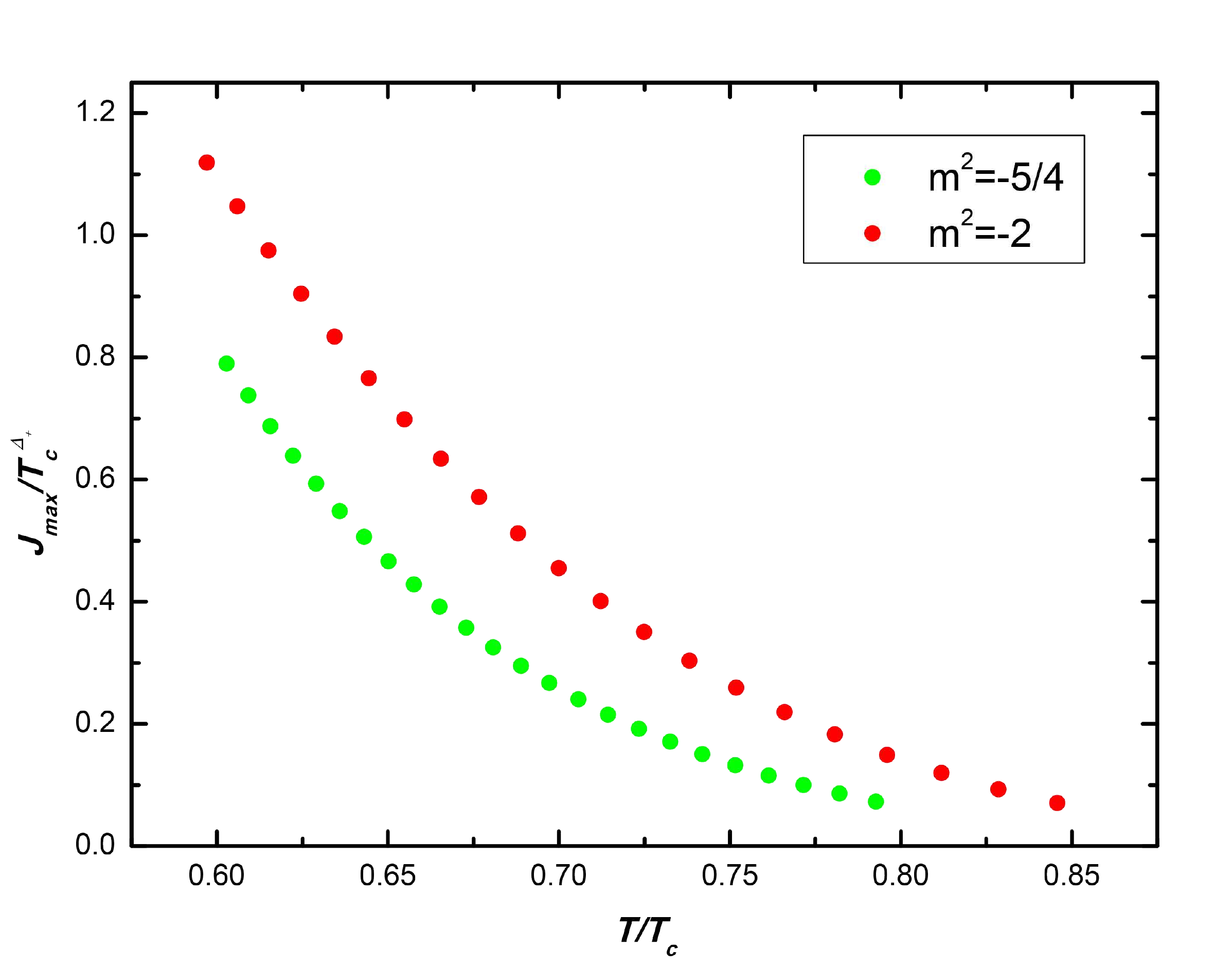}
  \end{center}
  \caption{The plot shows that the maximal current $J_{max}/T^{\Delta_{+}}_{c}$ goes down as $m^{2}$ grows. The value of $m^{2}=-2$(red), $-5/4$(green). The parameters are $\mu0=1$, $\mu_{\infty}=7.5$, $L=3$, $\epsilon=0.5$, $\sigma=0.5$ and $\kappa=1/100$.}
\label{JmaxTm}
\end{figure}

The relationship between the maximal current $J_{max}/T^{\Delta_{+}}_{c}$ and the temperature $T/T_{c}$ is shown in figure $\ref{JmaxTm}$. From the figure, we can see that $J_{max}/T^{\Delta_{+}}_{c}$ goes down as $m^{2}$ grows. The above results indicate that for fixed backreaction, the larger mass of the scalar field can hinder the scalar hair to form.

\section{(2+1)-Dimensional p-wave Josephson junction}\label{sec3}
\subsection{Holographic setup}
We consider the non-Abelian $SU(2)$ gauge filed in the (3+1)-dimensional Einstein gravity spacetime. The Lagrangian density is as follows
\begin{align}
\mathcal{L}&=R-2\Lambda-\frac{1}{4}F^{\mu\nu a}F_{\mu\nu}^{a},\label{pLagdensity}\\
F^{a}_{\mu\nu}&=\partial_{\mu}A_{\nu}^{a}-\partial_{\nu}A_{\mu}^{a}+g_{YM}\varepsilon^{abc}A_{\mu}^{b}A^{c}_{\nu}.\label{SU2strength}
\end{align}
Here, $F^{a}_{\mu\nu}$ is the strength of $SU(2)$ gauge field, and $a,b,c = (1,2,3)$ are the indices of the generators of $SU(2)$ algebra. The $A^{a}_{\mu}$ are the components of $A=A^{a}_{\mu}\tau^{a}dx^{\mu}$, where $\tau^{a}$ are the $SU(2)$ generators, which $[\tau^{a}, \tau^{b}] = \varepsilon^{abc}\tau^{c}$. The $\varepsilon^{abc}$ is the totally antisymmetric tensor. The coupling constant of the $SU(2)$ gauge field is $g_{YM}$.

By scaling the gauge field as $A\rightarrow A/g_{YM}$, Eqs.(\ref{pLagdensity}) and (\ref{SU2strength}) become
\begin{align}
\mathcal{L}&=R-2\Lambda+\kappa(-\frac{1}{4}\widetilde{F}^{\mu\nu a}\widetilde{F}_{\mu\nu}^{a}),\label{pLagdensity2}\\
\widetilde{F}^{a}_{\mu\nu}&=\partial_{\mu}A_{\nu}^{a}-\partial_{\nu}A_{\mu}^{a}+\varepsilon^{abc}A_{\mu}^{b}A^{c}_{\nu}.
\end{align}
Where $\kappa=1/g_{YM}^{2}$ is the strength of backreaction of the $SU(2)$ gauge field. We can also  see that the effect of backreaction decreases as $g_{YM}$ grows. The large limit ($g_{YM}\rightarrow\infty$) corresponds to the probe limit.

The equations of motion from the Lagrangian density ($\ref{pLagdensity}$) are
\begin{align}
\nabla_{\mu}F^{\mu\nu a}+\varepsilon^{abc}A_{\mu}^{b}F^{\mu\nu c}&=0,\label{su2equ}\\
R_{\mu\nu}+\frac{3}{\ell^{2}}g_{\mu\nu}-\kappa(\frac{1}{2}F^{a}_{\mu\lambda}F^{\lambda a}_{\nu}-\frac{1}{8}F^{a}_{\alpha\lambda}F^{\alpha\lambda a}g_{\mu\nu})&=0.\label{peinsteinequ}
\end{align}

In order to construct a holographic model of the p-wave Josephson junction, we bring in two bulk gauge fields $A_{1}^{3}=A_{z}$ and $A_{2}^{3}=A_{x}$ on basis of the model of p-wave superconductor. We consider the following ansatz of the $SU(2)$ gauge field
\begin{align}
A&=A_{0}^{3}\tau^{3}dt+A_{3}^{1}\tau^{1}dy+A_{1}^{3}\tau^{3}dz+A_{2}^{3}\tau^{3}dx\nonumber\\
 &=\phi\tau^{3}dt+\omega\tau^{1}dy+A_{z}\tau^{3}dz+A_{x}\tau^{3}dx,\label{pmatteransatz}
\end{align}
where $\phi$, $\omega$, $A_{z}$ and $A_{x}$ are all the real functions of coordinate $x$ and $z$. The EoMs ($\ref{su2equ}$) and ($\ref{peinsteinequ}$) are also a set of non-linear coupled partial equations with the matter ansatz ($\ref{pmatteransatz}$) and the metric ansatz ($\ref{metricansatz}$). These five equations derive from Eq.($\ref{su2equ}$), four of them are second-order PDEs and the remaining one is a first-order PDE which is   constrain equation. Seven equations from Eq.($\ref{peinsteinequ}$) are second-order PDEs with respect to $E_{i}$. Because each equation has thousands of terms, we can not write them down in paper. Obviously, we can only solve them numerically.

In the next step, we will study the following asymptotic forms of the matter field on the AdS boundary ($z=0$)
\begin{subequations}
\begin{align}
\omega(z,x)&\rightarrow z\omega^{(1)}(x)+z^{2}\omega^{(2)}(x)+\mathcal{O}(z^{3}),\\
\phi(z,x)&\rightarrow \mu(x)-z\rho(x)+\mathcal{O}(z^{2}),\\
A_{z}(z,x)&\rightarrow \mathcal{O}(z^{3}),\\
A_{x}(z,x)&\rightarrow \nu(x)+Jz+\mathcal{O}(z^{2}).
\end{align}
\end{subequations}
Here, we set $\omega^{(1)}=0$ and interpret $\langle\mathcal{O}\rangle=\omega^{(2)}$ as the condensation of p-wave superconductor.   The asymptotic behaviors of $E_{i}$ take the same forms as ($\ref{e1}$)$-$($\ref{e7}$) on the AdS boundary. The chemical potential $\mu$ and the phase difference $\gamma$ take ($\ref{chempoten}$) and ($\ref{phase}$), respectively.

The critical temperatures of p-wave Josephson junction with different values of $\kappa$ are shown as below
\begin{align}
T_{c}&\approx0.0654\mu(\infty),\quad \quad \kappa=0,\\
T_{c}&\approx0.0613\mu(\infty),\quad \quad \kappa=1/15,\\
T_{c}&\approx0.0578\mu(\infty),\quad \quad \kappa=1/8.
\end{align}
From the above, we can see that the critical temperature $T_{c}$ drops with increasing strength of backreaction $\kappa$.

\subsection{The numerical results}
\begin{figure}
\centering
\subfigure[]{\includegraphics[width=0.49\linewidth]{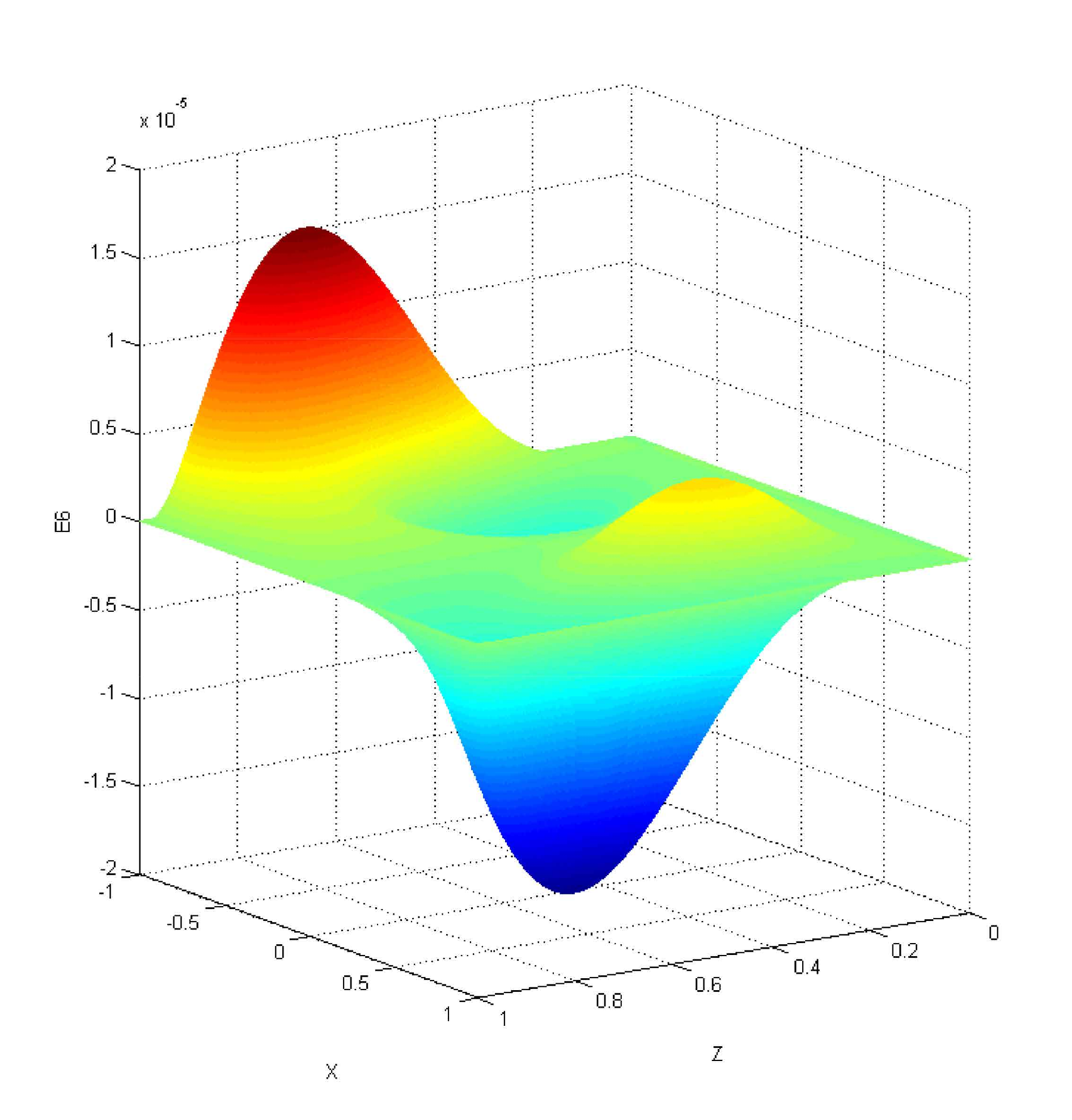}}
\hfill
\subfigure[]{\includegraphics[width=0.49\linewidth]{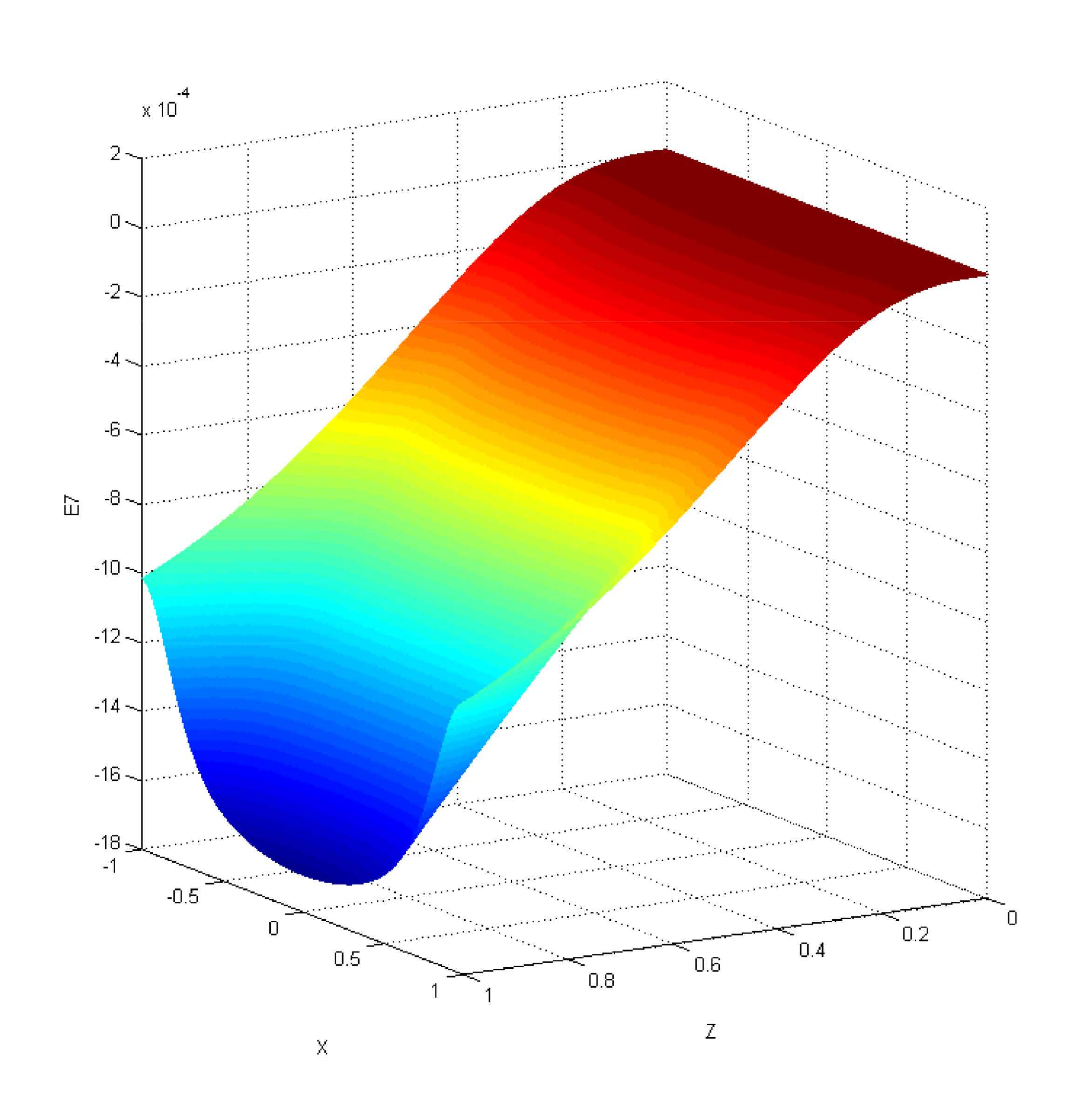}}
\caption{(a) The profile of $E_{6}$. (b) The profile of $E_{7}$. In all the plots, the parameters are $\mu0=1$, $\mu_{\infty}=5.5$, $L=3$, $\epsilon=0.6$, $\sigma=0.5$ and $\kappa=1/100$.}
\label{pwaveE6E7}
\end{figure}

In figure $\ref{pwaveE6E7}$ (a) and (b), we plot the profiles of the functions $E_{6}$ and $E_{7}$ . From these figures, we can see that $E_{6}$ is odd and $E_{7}$ is even .

\begin{figure}
  \begin{center}
  \includegraphics[width=0.9\textwidth,height=0.4\textheight]{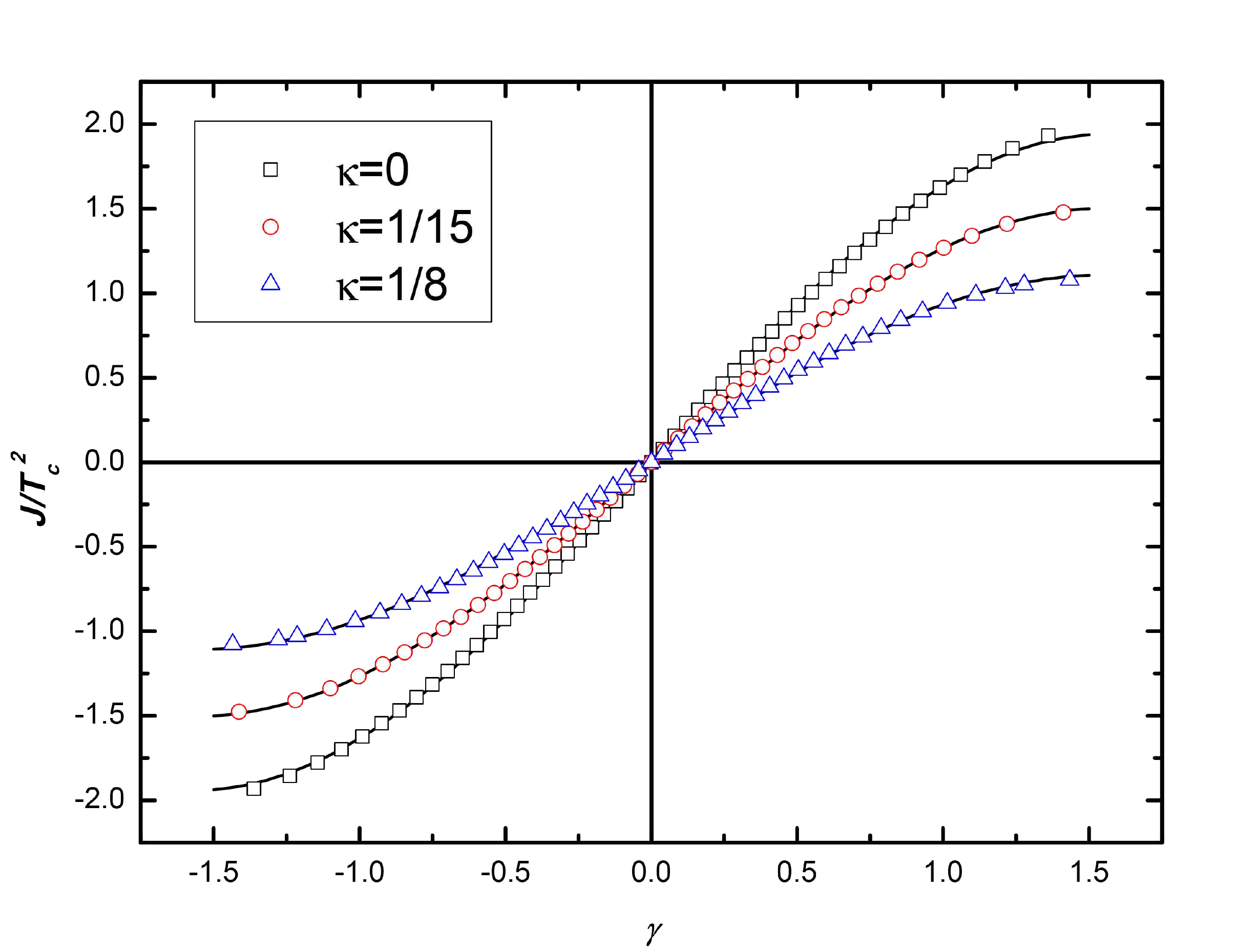}
  \end{center}
  \caption{The plot shows that the current $J/T_{c}^{2}$ is proportional to the phase difference $\gamma$. From top to bottom, curves correspond to $\kappa=0,1/15,1/8$. We use the parameters $\mu0=1$, $\mu_{\infty}=5.5$, $L=3$, $\epsilon=0.6$ and $\sigma=0.5$.}
\label{pwaveJgammak}
\end{figure}

We show the relationship between the current $J/T_{c}^{2}$ and the phase difference $\gamma$ in figure $\ref{pwaveJgammak}$. We take $\kappa=0,1/15,1/8$, respectively. The fitting results are
\begin{align}
J/T_{c}^{2}&\approx1.94186sin\gamma,\quad \quad \kappa=0,\\
J/T_{c}^{2}&\approx1.50461sin\gamma,\quad \quad \kappa=1/15,\\
J/T_{c}^{2}&\approx1.10894sin\gamma,\quad \quad \kappa=1/8.
\end{align}
From the above, we see that $J/T_{c}^{2}$ still varies as a sine function of $\gamma$ with fixing $\kappa$. Furthermore, $J/T_{c}^{2}$ will decrease when $\kappa$ grows.

\begin{figure}
\centering
\subfigure[]{\includegraphics[width=0.49\linewidth]{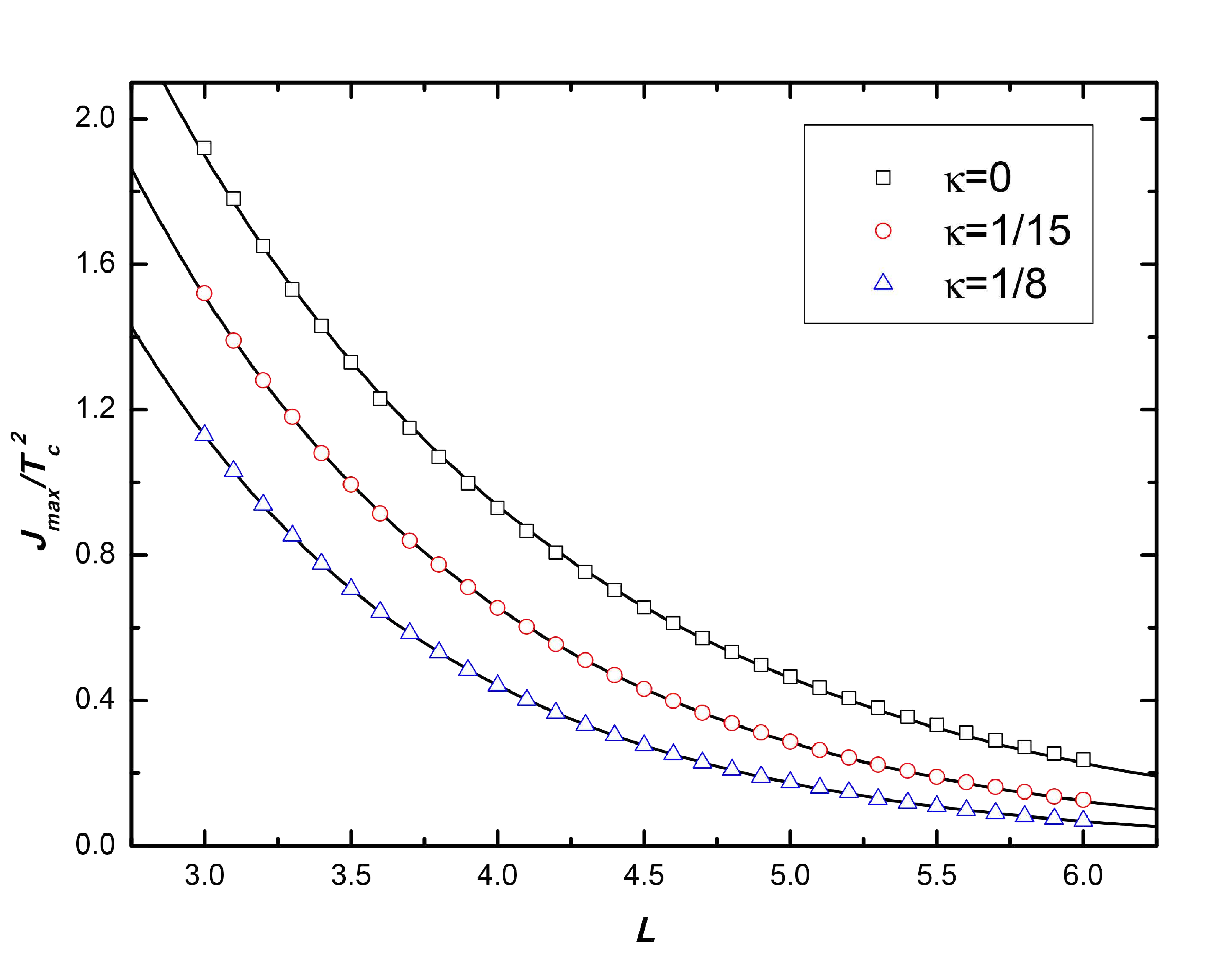}}
\hfill
\subfigure[]{\includegraphics[width=0.49\linewidth]{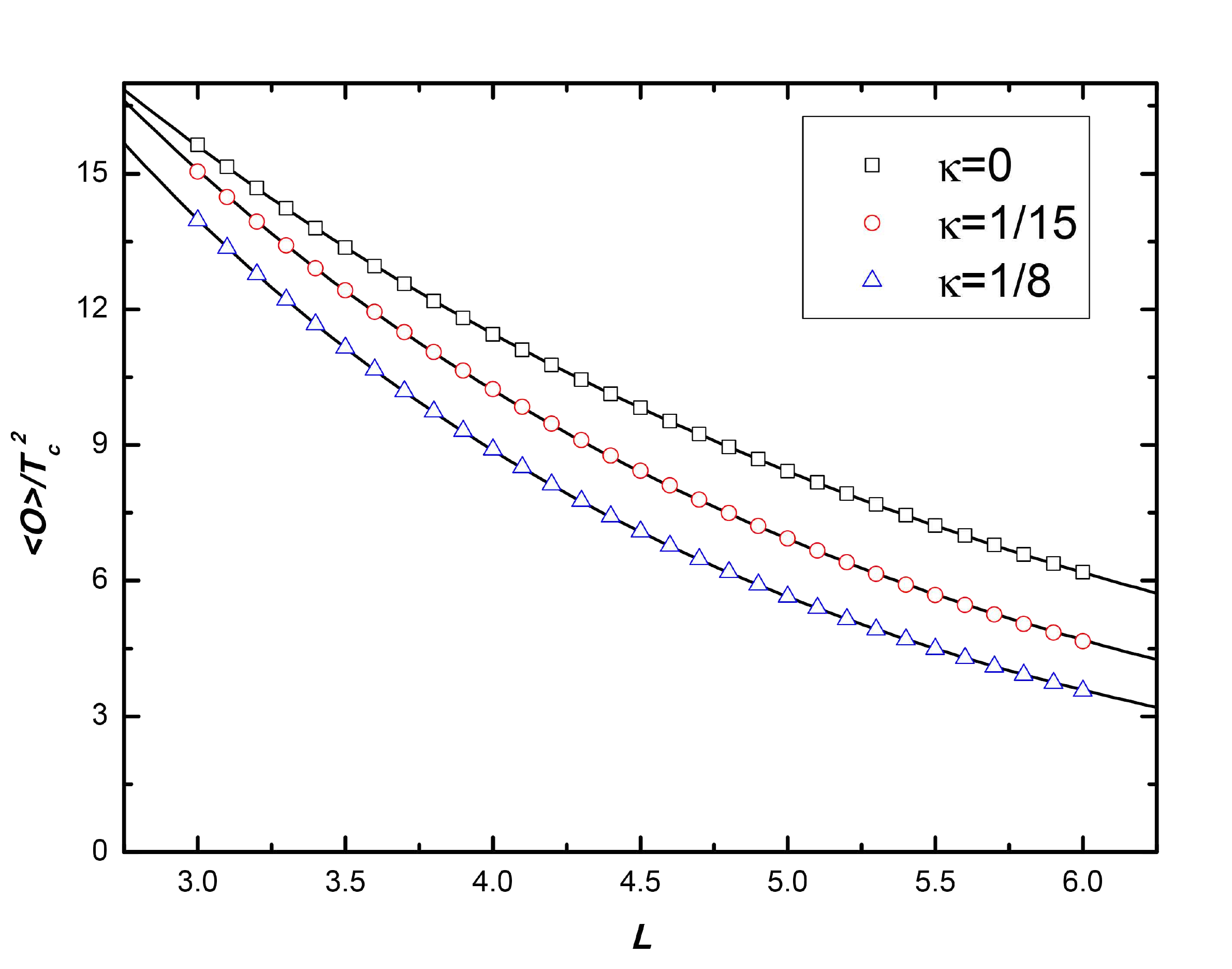}}
\caption{(a) The maximal current $J_{max}/T^{2}_{c}$ decreases exponentially with growing $L$. (b) The condensation $\langle O\rangle/T^{2}_{c}$ decreases exponentially with growing $L$. When $\kappa$ increases, $J_{max}/T^{2}_{c}$ and $\langle O\rangle/T^{2}_{c}$ will decrease, respectively. From top to bottom, curves correspond to $\kappa=0,1/15,1/8$. The parameters are $\mu0=1$, $\mu_{\infty}=5.5$, $\epsilon=0.6$ and $\sigma=0.5$.}
\label{pwaveJmaxLOLk}
\end{figure}

Figure $\ref{pwaveJmaxLOLk}$ (a) shows the relationship between the maximal current $J_{max}/T_{c}^{2}$ and the width of junction $L$ with increasing $\kappa=0,1/15,1/8$. The p-wave condensation $\langle O\rangle/T_{c}^{2}$ varies a function of $L$ with increasing $\kappa$ in figure $\ref{pwaveJmaxLOLk}$ (b). The fitting results are shown as follows
\begin{align}
J_{max}/T_{c}^{2}&\approx15.8458e^{-L/1.41421},\quad \quad \kappa=0,\\
J_{max}/T_{c}^{2}&\approx18.5319e^{-L/1.19729},\quad \quad \kappa=1/15,\\
J_{max}/T_{c}^{2}&\approx18.9418e^{-L/1.06426},\quad \quad \kappa=1/8.
\end{align}
and
\begin{align}
\langle O\rangle/T_{c}^{2}&\approx39.4108e^{-L/(2\times1.61943)},\quad \quad \kappa=0,\\
\langle O\rangle/T_{c}^{2}&\approx48.4891e^{-L/(2\times1.28419)},\quad \quad \kappa=1/15,\\
\langle O\rangle/T_{c}^{2}&\approx54.6341e^{-L/(2\times1.10103)},\quad \quad \kappa=1/8.
\end{align}
From the results, we know that by fixing $\kappa$, $J_{max}/T_{c}^{2}$ still decreases with growing $L$, exponentially. In addition, $J_{max}/T_{c}^{2}$ goes down as $\kappa$ increases. The relationship between $\langle O\rangle/T_{c}^{2}$ and $L$ is the same as the case of $J_{max}/T_{c}^{2}$.

\begin{figure}
  \begin{center}
  \includegraphics[width=0.9\textwidth,height=0.4\textheight]{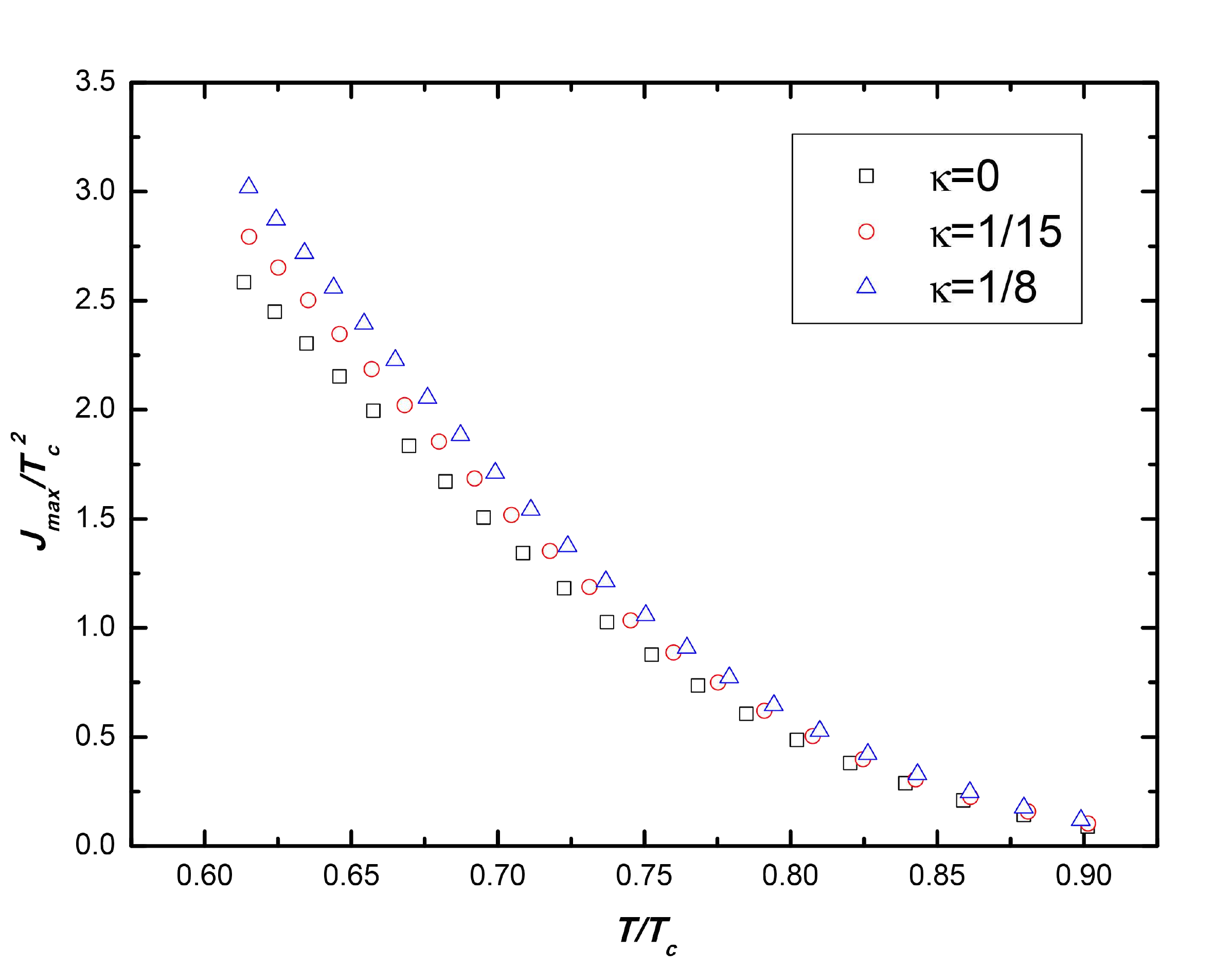}
  \end{center}
  \caption{The plot shows that $J_{max}/T_{c}^{2}$ decreases with increasing $T/T_{c}$ by fixing $\kappa$, and $J_{max}/T_{c}^{2}$ will go up, when $\kappa$ increases. From top to bottom, curves correspond to $\kappa=1/8,1/15,0$. The parameters are $\mu0=1$, $L=3$, $\epsilon=0.6$, $\sigma=0.5$.}
\label{pwaveJmaxTk}
\end{figure}

Finally, we represent the plot of between the maximal current $J_{max}/T_{c}^{2}$ and the temperature $T/T_{c}$ in figure $\ref{pwaveJmaxTk}$. From the figure, we know that $J_{max}/T_{c}^{2}$ decreases with growing $T/T_{c}$ by fixing $\kappa$, and $J_{max}/T_{c}^{2}$ goes down as $\kappa$ increases. The reasons that lead to the above results is the same as s-wave junction.

\section{Conclusion}\label{sec4}
Motivated by the effect of backreaction on the s-wave and p-wave holographic superconductors, we studied backreaction to the s-wave and p-wave Josephson junction, respectively. For s-wave Josephson junction, we  investigated the the full, dynamic equations of motion including Einstain equations in  the (3+1)-dimensional spacetime from the action of $U(1)$ gauge field and complex scalar field, and for p-wave Josephson junction, we considered the backreaction from the $SU(2)$ gauge field. We get a series of partial differential equations of fields that are nonlinear and coupled and solve them numerically. For the s-wave Josephson junction, we take two value of $m^{2}=-2,-5/4$. From the numerical results we can see that when the value of $m^{2}$ and the strength of backreaction $\kappa$ are fixed ($m^{2}=-2$), the current $J/T_{c}^{2}$ is still proportional to the sine of the phase difference $\gamma$. The reasons are that in the probe limit ($\kappa=0$), the spacetime and the matter (the sum of the gauge field and the scalar field) do not influence each other, the current is proportional to the sine of phase difference; when $\kappa\neq0$, the backreaction is turned on, note that the the spacetime affects the whole of the matter instead of the gauge field or the scalar field respectively, the relationship between the gauge field and the scalar field is not changed, so the current is still proportional to the sine of phase difference; in addition, the parameter $\kappa=1/q^{2}$ is related to the charge $q$, which is the coupling coefficient between the gauge field and the scalar field and represents the charge of Cooper pair in superconductor, when $\kappa$ varies, the charge $q$ will change, thus the backreaction only affects the amplitude of the current, so the relationship between current and phase difference holds for sine function with backreaction. The maximal current $J_{max}/T_{c}^{2}$ and the condensation $\langle O\rangle/T_{c}^{2}$ decreases exponentially with increasing the width of junction $L$, respectively, the maximal current $J_{max}/T_{c}^{2}$ decreases as the temperature $T/T_{c}$ grows. In addition, when the strength of backreaction $\kappa$ is fixed, $J/T_{c}^{\Delta_{+}}$, $\langle O\rangle/T_{c}^{\Delta_{+}}$ and $J_{max}/T_{c}^{\Delta_{+}}$ go down as $m^{2}$ increases. The reason is that the larger the mass of scalar field is, the harder the scalar hair forms. When the value of $m^{2}$ is fixed ($m^{2}=-2$), $J/T_{c}^{2}$, $\langle O\rangle/T_{c}^{2}$ and $J_{max}/T_{c}^{2}$ which varies with $L$ go down as $\kappa$ grows. The reasons are that the backreaction $\kappa$ makes the Cooper pairs generate harder, and the charge $q$ and the coherence length become smaller. It is noteworthy that $J_{max}/T_{c}^{2}$ which varies with $T/T_{c}$ goes up as $\kappa$ increases, this is because the effection of low temperature is stronger than that of backreaction. For p-wave Josephson junction, the property is the same as the case of s-wave Josephson junction.

Now, we have studied holographic models of s-N-s and p-N-p Josephson junctions away from probe limit, moreover we would like to investigate the s-I-s and p-I-p junctions with backreaction in the future. In\cite{Li:2014xia,Hu:2015dnl}, the s-N-s Josephson junction in the probe limit have been studied in the Lifshitz gravity and massive gravity, respectively. Thus our next study would be to generalize the above work to the case with backreaction .

\section*{Acknowledgement}
YQW would like to thank Rong-Gen Cai and Yu-Xiao Liu for very helpful discussion. SL and YQW were supported by the National Natural Science Foundation of China.


\begin{thebibliography}{99}
\bibitem{Maldacena:1997re}
  J.~M.~Maldacena,
  {\em The large $N$ limit of superconformal field theories and supergravity}.
  Int.\ J.\ Theor.\ Phys.\  {\bf 38}, 1113 (1999).

\bibitem{Maldacena:1998re}
  J.~M.~Maldacena,
  Adv.Theor.\ Math.\ Phys.\  {\bf 2}, 231 (1998)
  [hep-th/9711200].

\bibitem{Witten:1998qj}
  E.~Witten,
 {\em Anti-de Sitter space and holography}.
  Adv.\ Theor.\ Math.\ Phys.\  {\bf 2}, 253 (1998)
  [hep-th/9802150].

\bibitem{Aharony:1999ti}
  O.~Aharony, S.~S.~Gubser, J.~M.~Maldacena, H.~Ooguri and Y.~Oz,
 {\em Large N field theories, string theory and gravity}.
  Phys.\ Rept.\  {\bf 323}, 183 (2000)
  [hep-th/9905111].

\bibitem{Gubser:2008px}
  S.~S.~Gubser,
  {\em Breaking an Abelian gauge symmetry near a black hole horizon}.
  Phys.\ Rev.\ D {\bf 78}, 065034 (2008)
  [arXiv:0801.2977 [hep-th]].

\bibitem{Hartnoll:2008vx}
  S.~A.~Hartnoll, C.~P.~Herzog and G.~T.~Horowitz,
  {\em Building a holographic superconductor}.
  Phys.\ Rev.\ Lett.\  {\bf 101}, 031601 (2008)
  [arXiv:0803.3295 [hep-th]].

\bibitem{Gubser:2008wv}
  S.~S.~Gubser and S.~S.~Pufu,
  {\em The Gravity dual of a p-wave superconductor}.
  JHEP {\bf 0811}, 033 (2008)
  [arXiv:0805.2960 [hep-th]].

\bibitem{Chen:2010mk}
  J.~W.~Chen, Y.~J.~Kao, D.~Maity, W.~Y.~Wen and C.~P.~Yeh,
  {\em Towards A Holographic Model of D-Wave Superconductors}.
  Phys.\ Rev.\ D {\bf 81}, 106008 (2010)
  [arXiv:1003.2991 [hep-th]].

\bibitem{Benini:2010pr}
  F.~Benini, C.~P.~Herzog, R.~Rahman and A.~Yarom,
  {\em Gauge gravity duality for d-wave superconductors: prospects and challenges}.
  JHEP {\bf 1011}, 137 (2010)
  [arXiv:1007.1981 [hep-th]].


\bibitem{Basu:2010fa}
  P.~Basu, J.~He, A.~Mukherjee, M.~Rozali and H.~H.~Shieh,
  {\em Competing holographic orders}.
  JHEP {\bf 1010}, 092 (2010)
  [arXiv:1007.3480 [hep-th]].

\bibitem{Musso:2013rnr}
  D.~Musso,
  {\em Competition/Enhancement of Two Probe Order Parameters in the Unbalanced Holographic Superconductor}.
  JHEP {\bf 1306}, 083 (2013)
  [arXiv:1302.7205 [hep-th]].

\bibitem{Cai:2013wma}
  R.~G.~Cai, L.~Li, L.~F.~Li and Y.~Q.~Wang,
  {\em Competition and coexistence of order parameters in holographic multi-band superconductors}.
  JHEP {\bf 1309}, 074 (2013)
  [arXiv:1307.2768 [hep-th]].

\bibitem{Amoretti:2013oia}
  A.~Amoretti, A.~Braggio, N.~Maggiore, N.~Magnoli and D.~Musso,
  {\em Coexistence of two vector order parameters: a holographic model for ferromagnetic superconductivity}.
  JHEP {\bf 1401}, 054 (2014)
  [arXiv:1309.5093 [hep-th]].

\bibitem{Li:2014wca}
  L.~F.~Li, R.~G.~Cai, L.~Li and Y.~Q.~Wang,
  {\em Competition between s-wave order and d-wave order in holographic superconductors}.
  JHEP {\bf 1408}, 164 (2014)
  [arXiv:1405.0382 [hep-th]].

\bibitem{Nie:2013sda}
  Z.~Y.~Nie, R.~G.~Cai, X.~Gao and H.~Zeng,
  {\em Competition between the s-wave and p-wave superconductivity phases in a holographic model}.
  JHEP {\bf 1311}, 087 (2013)
  [arXiv:1309.2204 [hep-th]].

\bibitem{Amado:2013lia}
  I.~Amado, D.~Arean, A.~Jimenez-Alba, L.~Melgar, I.~Salazar Landea,
  {\em Holographic s+p Superconductors}.
  Phys.\ Rev.\ D {\bf 89}, no. 2, 026009 (2014)
  [arXiv:1309.5086 [hep-th]].



\bibitem{Nie:2014qma}
  Z.~Y.~Nie, R.~G.~Cai, X.~Gao, L.~Li and H.~Zeng,
  {\em Phase transitions in a holographic s  $+$  p model with back-reaction}.
  Eur.\ Phys.\ J.\ C {\bf 75}, 559 (2015)
  [arXiv:1501.00004 [hep-th]].


\bibitem{Hartnoll:2009sz}
  S.~A.~Hartnoll,
  {\em Lectures on holographic methods for condensed matter physics}.
  Class.\ Quant.\ Grav.\  {\bf 26}, 224002 (2009)
  [arXiv:0903.3246 [hep-th]].

\bibitem{Herzog:2009xv}
  C.~P.~Herzog,
  {\em Lectures on holographic superfluidity and superconductivity}.
  J.\ Phys.\ A {\bf 42}, 343001 (2009)
  [arXiv:0904.1975 [hep-th]].

\bibitem{Horowitz:2010gk}
  G.~T.~Horowitz,
  {\em Introduction to holographic superconductors}.
  Lect.\ Notes Phys.\  {\bf 828}, 313 (2011)
  [arXiv:1002.1722 [hep-th]].

\bibitem{Cai:2015cya}
  R.~G.~Cai, L.~Li, L.~F.~Li and R.~Q.~Yang,
  {\em Introduction to Holographic Superconductor Models}.
  Sci.\ China Phys.\ Mech.\ Astron.\  {\bf 58}, no. 6, 060401 (2015)
  [arXiv:1502.00437 [hep-th]].





\bibitem{Hartnoll:2008kx}
  S.~A.~Hartnoll, C.~P.~Herzog and G.~T.~Horowitz,
  {\em Holographic superconductors}.
  JHEP {\bf 0812}, 015 (2008)
  [arXiv:0810.1563 [hep-th]].










\bibitem{Ammon:2009xh}
  M.~Ammon, J.~Erdmenger, V.~Grass, P.~Kerner and A.~O'Bannon,
  {\em On Holographic p-wave Superfluids with Back-reaction}.
  Phys.\ Lett.\ B {\bf 686}, 192 (2010)
  [arXiv:0912.3515 [hep-th]].

\bibitem{Brihaye:2010mr}
  Y.~Brihaye and B.~Hartmann,
  {\em Holographic Superconductors in 3+1 dimensions away from the probe limit}.
  Phys.\ Rev.\ D {\bf 81}, 126008 (2010)
  [arXiv:1003.5130 [hep-th]].

\bibitem{Cai:2010zm}
  R.~G.~Cai, Z.~Y.~Nie and H.~Q.~Zhang,
  {\em Holographic Phase Transitions of P-wave Superconductors in Gauss-Bonnet Gravity with Back-reaction}.
  Phys.\ Rev.\ D {\bf 83}, 066013 (2011)
  [arXiv:1012.5559 [hep-th]].

\bibitem{Pan:2011ns}
  Q.~Pan and B.~Wang,
  {\em General holographic superconductor models with backreactions}.
  [arXiv:1101.0222 [hep-th]].






\bibitem{Liu:2011fy}
  Y.~Liu, Q.~Pan and B.~Wang,
  {\em Holographic superconductor developed in BTZ black hole background with backreactions}.
  Phys.\ Lett.\ B {\bf 702}, 94 (2011)
  [arXiv:1106.4353 [hep-th]].

\bibitem{Peng:2012vb}
  Y.~Peng, X.~M.~Kuang, Y.~Liu and B.~Wang,
  {\em Phase transition in the holographic model of superfluidity with backreactions}.
  [arXiv:1204.2853 [hep-th]].

\bibitem{Pan:2012jf}
  Q.~Pan, J.~Jing, B.~Wang and S.~Chen,
  {\em Analytical study on holographic superconductors with backreactions}.
  JHEP {\bf 1206}, 087 (2012)
  [arXiv:1205.3543 [hep-th]].

\bibitem{Ge:2012vp}
  X.~H.~Ge, S.~F.~Tu and B.~Wang,
  {\em d-Wave holographic superconductors with backreaction in external magnetic fields}.
  JHEP {\bf 1209}, 088 (2012)
  [arXiv:1209.4272 [hep-th]].

\bibitem{Arias:2012py}
  R.~E.~Arias and I.~S.~Landea,
  {\em Backreacting p-wave Superconductors}.
  JHEP {\bf 1301}, 157 (2013)
  [arXiv:1210.6823 [hep-th]].

\bibitem{Nakonieczny:2014pma}
  ~Nakonieczny and M.~Rogatko,
  {\em Analytic study on backreacting holographic superconductors with dark matter sector}.
  Phys.\ Rev.\ D {\bf 90}, no. 10, 106004 (2014)
  [arXiv:1411.0798 [hep-th]].

\bibitem{Horowitz:2012ky}
  G.~T.~Horowitz, J.~E.~Santos and D.~Tong,
  {\em Optical Conductivity with Holographic Lattices}.
  JHEP {\bf 1207}, 168 (2012)
  [arXiv:1204.0519 [hep-th]].

\bibitem{Horowitz:2012gs}
  G.~T.~Horowitz, J.~E.~Santos and D.~Tong,
  {\em Further Evidence for Lattice-Induced Scaling}.
  JHEP {\bf 1211}, 102 (2012)
  [arXiv:1209.1098 [hep-th]].

\bibitem{Horowitz:2013jaa}
  G.~T.~Horowitz and J.~E.~Santos,
  {\em General Relativity and the Cuprates}.
  JHEP {\bf 1306}, 087 (2013)
  [arXiv:1302.6586 [hep-th]].

\bibitem{Ling:2013aya}
  Y.~Ling, C.~Niu, J.~P.~Wu, Z.~Y.~Xian and H.~b.~Zhang,
  {\em Holographic Fermionic Liquid with Lattices}.
  JHEP {\bf 1307}, 045 (2013)
  [arXiv:1304.2128 [hep-th]].

\bibitem{Bao:2013ixa}
  N.~Bao and S.~Harrison,
  {\em Crystalline Scaling Geometries from Vortex Lattices}.
  Phys.\ Rev.\ D {\bf 88}, 046009 (2013)
  [arXiv:1306.1532 [hep-th]].

\bibitem{Mozaffar:2013bva}
  M.~R.~M.~Mozaffar and A.~Mollabashi,
  {\em Crystalline geometries from a fermionic vortex lattice}.
  Phys.\ Rev.\ D {\bf 89}, no. 4, 046007 (2014)
  [arXiv:1307.7397 [hep-th]].

\bibitem{Ishibashi:2013nsa}
  A.~Ishibashi and K.~Maeda,
  {\em Thermalization of boosted charged AdS black holes by an ionic Lattice}.
  Phys.\ Rev.\ D {\bf 88}, no. 6, 066009 (2013)
  [arXiv:1308.5740 [hep-th]].

\bibitem{Donos:2013eha}
  A.~Donos and J.~P.~Gauntlett,
  {\em Holographic Q-lattices}.
  JHEP {\bf 1404}, 040 (2014)
  [arXiv:1311.3292 [hep-th]].

\bibitem{Iizuka:2013wya}
  N.~Iizuka, A.~Ishibashi and K.~Maeda,
  {\em Persistent Superconductor Currents in Holographic Lattices}.
  Phys.\ Rev.\ Lett.\  {\bf 113}, 011601 (2014)
  [arXiv:1312.6124 [hep-th]].

\bibitem{Aprile:2014aja}
  F.~Aprile and T.~Ishii,
  {\em A Simple Holographic Model of a Charged Lattice}.
  JHEP {\bf 1410}, 151 (2014)
  [arXiv:1406.7193 [hep-th]].

\bibitem{Donos:2014yya}
  A.~Donos and J.~P.~Gauntlett,
  {\em The thermoelectric properties of inhomogeneous holographic lattices}.
  JHEP {\bf 1501}, 035 (2015)
  [arXiv:1409.6875 [hep-th]].

\bibitem{Ling:2014laa}
  Y.~Ling, P.~Liu, C.~Niu, J.~P.~Wu and Z.~Y.~Xian,
  {\em Holographic Superconductor on Q-lattice}.
  JHEP {\bf 1502}, 059 (2015)
  [arXiv:1410.6761 [hep-th]].

\bibitem{Ling:2014bda}
  Y.~Ling, P.~Liu, C.~Niu, J.~P.~Wu and Z.~Y.~Xian,
  {\em Holographic fermionic system with dipole coupling on Q-lattice}.
  JHEP {\bf 1412}, 149 (2014)
  [arXiv:1410.7323 [hep-th]].

\bibitem{Chen:2015azo}
  L.~K.~Chen, H.~Guo and F.~W.~Shu,
  {\em Crystalline geometries from fermionic vortex lattice with hyperscaling violation}.
  [arXiv:1511.01370 [hep-th]].



\bibitem{Andrade:2015iyf}
  T.~Andrade and A.~Krikun,
  {\em Commensurability effects in holographic homogeneous lattices}.
  JHEP {\bf 1605}, 039 (2016)
  [arXiv:1512.02465 [hep-th]].





\bibitem{Alsup:2016fii}
  J.~Alsup, E.~Papantonopoulos, G.~Siopsis and K.~Yeter,
  {\em Holographic Fermi Liquids in a Spontaneously Generated Lattice}.
  Phys.\ Rev.\ D {\bf 93}, no. 10, 105045 (2016)
  [arXiv:1603.03382 [hep-th]].

\bibitem{Josephson:1962zz}
  B.~D.~Josephson,
  {\em  Possible new effects in superconductive tunnelling}.
  Phys.\ Lett.\  {\bf 1}, 251 (1962).

\bibitem{Horowitz:2011dz}
  G.~T.~Horowitz, J.~E.~Santos and B.~Way,
  {\em A holographic Josephson junction}.
  Phys.\ Rev.\ Lett.\  {\bf 106}, 221601 (2011)
  [arXiv:1101.3326 [hep-th]].

\bibitem{Wang:2011rva}
  Y.~Q.~Wang, Y.~X.~Liu and Z.~H.~Zhao,
  {\em Holographic Josephson junction in 3+1 dimensions}.
  [arXiv:1104.4303 [hep-th]].

\bibitem{Siani:2011uj}
  M.~Siani,
  {\em On inhomogeneous holographic superconductors}.
  [arXiv:1104.4463 [hep-th]].

\bibitem{Kiritsis:2011zq}
  E.~Kiritsis and V.~Niarchos,
  {\em Josephson junctions and AdS/CFT networks}.
  JHEP {\bf 1107}, 112 (2011)
  [Erratum-ibid.\  {\bf 1110}, 095 (2011)]
  [arXiv:1105.6100 [hep-th]].

\bibitem{Wang:2011ri}
  Y.~Q.~Wang, Y.~X.~Liu and Z.~H.~Zhao,
  {\em Holographic p-wave Josephson junction}.
  [arXiv:1109.4426 [hep-th]].

\bibitem{Wang:2012yj}
  Y.~Q.~Wang, Y.~X.~Liu, R.~G.~Cai, S.~Takeuchi and H.~Q.~Zhang,
  {\em Holographic SIS Josephson junction}.
  JHEP {\bf 1209}, 058 (2012)
  [arXiv:1205.4406 [hep-th]].

\bibitem{Cai:2013sua}
  R.~G.~Cai, Y.~Q.~Wang and H.~Q.~Zhang,
  {\em A holographic model of SQUID}.
  JHEP {\bf 1401}, 039 (2014)
  [arXiv:1308.5088 [hep-th]].

\bibitem{Takeuchi:2013kra}
  S.~Takeuchi,
  {\em Holographic Superconducting Quantum Interference Device}.
  Int.\ J.\ Mod.\ Phys.\ A {\bf 30}, no. 09, 1550040 (2015)
  [arXiv:1309.5641 [hep-th]].

\bibitem{Li:2014xia}
  H.~F.~Li, L.~Li, Y.~Q.~Wang and H.~Q.~Zhang,
  {\em Non-relativistic Josephson Junction from Holography}.
  JHEP {\bf 1412}, 099 (2014)
  [arXiv:1410.5578 [hep-th]].

\bibitem{Liu:2015zca}
  S.~Liu and Y.~Q.~Wang,
  {\em Holographic model of hybrid and coexisting s-wave and p-wave Josephson junction}.
  Eur.\ Phys.\ J.\ C {\bf 75}, no. 10, 493 (2015)
  [arXiv:1504.06918 [hep-th]].

\bibitem{Hu:2015dnl}
  Y.~P.~Hu, H.~F.~Li, H.~B.~Zeng and H.~Q.~Zhang,
  {\em Holographic Josephson Junction from Massive Gravity}.
  Phys.\ Rev.\ D {\bf 93}, no. 10, 104009 (2016)
  [arXiv:1512.07035 [hep-th]].

\bibitem{Breitenlohner:1982bm}
  P.~Breitenlohner and D.~Z.~Freedman,
  {\em Positive energy in anti-De Sitter backgrounds and gauged extended
  supergravity}.
  Phys.\ Lett.\  {\bf B 115}, 197 (1982).

\bibitem{Basu:2008st}
  P.~Basu, A.~Mukherjee and H.~-H.~Shieh,
  {\em Supercurrent: vector hair for an AdS black hole}.
  Phys.\ Rev.\  {\bf D 79}, 045010 (2009)
  [arXiv:0809.4494 [hep-th]].

\bibitem{Herzog:2008he}
  C.~P.~Herzog, P.~K.~Kovtun and D.~T.~Son,
  {\em Holographic model of superfluidity}.
  Phys.\ Rev.\  {\bf D 79}, 066002 (2009)
  [arXiv:0809.4870 [hep-th]].

\bibitem{Arean:2010xd}
  D.~Arean, M.~Bertolini, J.~Evslin and T.~Prochazka,
  {\em On holographic superconductors with DC current}.
  JHEP {\bf 1007}, 060 (2010)
  [arXiv:1003.5661 [hep-th]].

\bibitem{Sonner:2010yx}
  J.~Sonner and B.~Withers,
  {\em A gravity derivation of the Tisza$-$Landau model in AdS/CFT}.
  Phys.\ Rev.\  {\bf D 82}, 026001 (2010)
  [arXiv:1004.2707 [hep-th]].

\bibitem{Horowitz:2008bn}
  G.~T.~Horowitz and M.~M.~Roberts,
  {\em Holographic superconductors with various condensates}.
  Phys.\ Rev.\  {\bf D 78}, 126008 (2008)
  [arXiv:0810.1077 [hep-th]].

\bibitem{Arean:2010zw}
  D.~Arean, P.~Basu and C.~Krishnan,
  {\em The many phases of holographic superfluids}.
  JHEP {\bf 1010}, 006 (2010)
  [arXiv:1006.5165 [hep-th]].

\bibitem{Zeng:2010fs}
  H.~B.~Zeng, W.~M.~Sun and H.~S.~Zong,
  {\em Supercurrent in p-wave holographic superconductor}.
  Phys.\ Rev.\ D {\bf 83}, 046010 (2011)
  [arXiv:1010.5039 [hep-th]].

\bibitem{Arean:2010wu}
  D.~Arean, M.~Bertolini, C.~Krishnan and T.~Prochazka,
  {\em Type IIB Holographic Superfluid Flows}.
  JHEP {\bf 1103}, 008 (2011)
  [arXiv:1010.5777 [hep-th]].

















\end{thebibliography}
\end{document}